\documentclass[12pt]{article}

\usepackage{amsmath,amssymb,graphicx,multicol} 
\usepackage{pstricks}
\usepackage{cite}
\usepackage{epsfig}

\newcommand{\ud}{\,\mathrm{d}}

\newcommand{\pd}{\partial}

\definecolor{orange}{rgb}{1, 0.4, 0} 
\definecolor{vertfonce}{rgb}{0, 0.4, 0} 
\definecolor{marron}{rgb}{0.36,0.13,0.00} 
\definecolor{purple}{rgb}{0.4,0.0,0.4} 
\definecolor{pink}{rgb}{0.8,0.3,0.6} 
\definecolor{gray}{rgb}{0.3,0.3,0.3} 

\newcommand{\beq}{\begin{eqnarray}}
\newcommand{\eeq}{\end{eqnarray}}

\newcommand{\centeron}[2]{{\setbox0=\hbox{#1}\setbox1=\hbox{#2}\ifdim

\wd1>\wd0\kern.5\wd1\kern-.5\wd0\fi
\copy0

\kern-.5\wd0\kern-.5\wd1\copy1\ifdim\wd0>\wd1
                                     \kern.5\wd0\kern-.5\wd1\fi}}
\newcommand{\ltap}{\>\centeron{\raise.35ex\hbox{$<$}}
                             {\lower.65ex\hbox{$\sim$}}\>}
\newcommand{\gtap}{\>\centeron{\raise.35ex\hbox{$>$}}
                             {\lower.65ex\hbox{$\sim$}}\>}

\newcommand\ZZ{\hbox{\zfont Z\kern-.4emZ}}
\font\zfont = cmss10 

\newcommand{\kappagamma}{\kappa_{\gamma \gamma}}
\newcommand{\kappaglu}{\kappa_{gg}}

\begin{document}
\begin{titlepage}
\begin{flushright}
LYCEN 2008-13 
\end{flushright}

\vskip.5cm
\begin{center}
{\huge \bf 
$H \to \gamma \gamma$ beyond the Standard Model}

\vskip.1cm
\end{center}
\vskip1cm

\begin{center}
{\bf
{Giacomo Cacciapaglia}, {Aldo Deandrea}, and {J\'er\'emie Llodra-Perez}}
\end{center}
\vskip 8pt

\begin{center}
{\it
Universit\'e de Lyon, F-69622 Lyon, France; Universit\'e Lyon 1, Villeurbanne;\\
CNRS/IN2P3, UMR5822, Institut de Physique Nucl\'eaire de Lyon\\
F-69622 Villeurbanne Cedex, France} \\
\end{center}

\vskip2truecm

\begin{abstract}
\vskip 3pt
\noindent
We consider the $H \to \gamma\gamma$ decay process and the gluon fusion production of a light Higgs, and provide a general framework for testing models of new physics beyond the Standard Model. 
We apply our parametrisation to typical models extending the Standard Model in 4 and 5 dimensions, and show how the parametrisation can be used to discriminate between different scenarios of new physics at the Large Hadron Collider and at future Linear Colliders.

\end{abstract}

\end{titlepage}

\newpage


\setcounter{footnote}{0}
\section{Introduction}
\label{sec:intro}
\setcounter{equation}{0}

The decay of the Higgs in two photons is one of the most important discovery channels at the Large Hadron Collider (LHC), and it is certainly the golden mode at low masses, where the decay channels into heavy gauge bosons are closed.
Detailed studies, including detector simulations, in the Standard Model (SM) and in its supersymmetric extensions are available~\cite{atlascms}.
This mode is also a powerful probe of the electroweak symmetry breaking sector of the theory, because it is a loop-induced process, therefore it is sensitive to any particle with a large coupling to the Higgs.
In the SM it depends primarily on the couplings of the Higgs boson with heavy quarks (the top) and gauge bosons (the $W$), whose masses are tightly related to the electroweak scale.
In any extension of the SM, particles that do couple strongly to the Higgs, and therefore play a role in the breaking of the electroweak symmetry, will also contribute to this loop and modify the SM prediction.
For instance, new particles at the TeV scale are required to soften the divergences that appear in the corrections to the Higgs mass generated by top and $W$-$Z$ loops.
Many models in fact predict the existence of partners of the top and $W$: stops and gauginos in supersymmetry, heavy $W$'s and tops in extra dimensional models and Little Higgs models, and so on.
Studying this channel will therefore give an indirect access to the mechanism underlying the electroweak symmetry breaking.
At the LHC, we also need to take into account the Higgs production mechanism. 
In the range of masses below the $H \to WW$ decay threshold (we shall consider the range 115 GeV to 150 GeV), the largest production cross section for the Higgs boson in the SM is via gluon fusion ($gg \to H$) which, in the SM, is roughly an order of magnitude larger than vector boson fusion and other processes. 
While some of the production channels may have additional leptons, jets or missing energy in their final state, it will be difficult, at least at low luminosity, to take advantage of these different signatures. We shall therefore consider mainly the inclusive $H \to \gamma\gamma$ process. The interest of performing exclusive studies like the production via vector boson fusion, will be also discussed as it allows to better discriminate the kind of new physics that can be tested in the $H \to \gamma\gamma$ mode.
However large integrated luminosity is necessary in this case.
The main production process $gg \to H$ is a loop induced process like the decay $H \to \gamma\gamma$, and it is sensitive to the same particles and physics.

In this paper we study the photon channel with the purpose of performing a model independent analysis, allowing to determine the possibility and the limits for discriminating various scenarios of new physics. 
In the following we shall propose a model independent parametrisation of these loop processes in order to test the possibility of discrimination of various models of new physics. 
We shall provide a general and simple formalism to easily calculate the contribution of the new heavy states given their spectrum.
We will assume that the new physics only affects those two processes, and corrections to the other production and decay channels are ignored.
In the SM it is well known that the contribution of heavy particles to $H \to \gamma\gamma$ and $H \to gg$ processes does not decouple for particle masses much larger than the Higgs boson one. 
The reason is that these SM masses are uniquely generated by the coupling to the Higgs boson and the mass dependence of their coupling cancels the mass dependence in the loop integral. 
In general extensions of the SM this is not necessarily the case, as the masses may receive other contributions.
The effect on the decay can therefore be sensitive to the mass scale of the new physics.
Studying this channel in detail can give some hints about the model of new physics, and this information will be complementary to the direct discovery of new states at the LHC.
Finally, the precise determination of the Higgs branching ratios at future Linear Collider will be an even more powerful discrimination tool, even when the new particles are well beyond the direct production threshold at the Linear Collider. 

In the next section we settle our notation and define our parametrisation of the loop induced processes $H \to \gamma\gamma$ and $H \to gg$. 
In the sections \ref{sec:models4} and \ref{sec:models5} we consider various scenarios of new physics in 4 and 5 dimensions, in section \ref{sec:numres} we discuss numerical results in various models and how the parametrisation we propose can provide a hint to what kind of new physics can be deduced from data both at the LHC and at Linear Colliders.
Finally we give our conclusions, and we leave details on the calculation to the appendices.

\section{Definitions and notations}
\label{sec:definitions}

In order to establish our notations, we will briefly review the decay of the Higgs in photons and gluons (the decay width in gluons is directly related to the gluon-fusion production cross section at hadronic colliders). 
The decay widths can be written as:
\beq
\Gamma_{\gamma \gamma} &=& \frac{G_F \alpha^2 m_H^3}{128 \sqrt{2} \pi^3} \left| A_W (\tau_W) + \sum_{\rm fermions} N_{c,f} Q_f^2 A_F (\tau_f) + \sum_{NP} N_{c,NP} Q_{NP}^2 A_{NP} (\tau_{NP}) \right|^2\,, \\
\Gamma_{g g} &=& \frac{G_F \alpha_s^2 m_H^3}{16 \sqrt{2} \pi^3} \left| \frac{1}{2} \sum_{\rm quarks} A_F (\tau_f) +  \sum_{NP} C (r_{NP})  A_{NP} (\tau_{NP}) \right|^2\,,
\eeq
where $\tau_x = \frac{m_H^2}{4 m_x^2}$, $N_{c,x}$ is the number of colour states in the colour representation (3 for quarks, 1 for leptons), the constant $C (r)$ is an SU(3) colour factor (defined as ${\rm Tr}[ t^a_r t^b_r] = C(r) \delta^{ab}$ where $t_r^a$ are the SU(3) generators in the representation $r$; it is equal to $1/2$ for the quarks and $3$ for an adjoint), $Q_x$ is the electric charge of the particle in the loop, and the functions $A (\tau)$ depend on the spin and couplings to the Higgs of the particle running in the loop.

In the SM, all masses are proportional to the Higgs vacuum expectation value (VEV) $v$, therefore the couplings to the Higgs can be written as
\beq
y_{h \bar f f}^{SM} &=& \frac{m_f}{v} \quad \mbox{for fermions}\,, \\
y_{h W W}^{SM} &=& 2 \frac{m_W^2}{v} \quad \mbox{for bosons}\,.
\eeq
Under this assumption, the amplitudes are given by ($F$ stands for spin-$1/2$ fermions, $W$ for vector bosons and $S$ for scalars)~\cite{Spira}
\beq
A_F (\tau) &=& \frac{2}{\tau^2} \left( \tau + (\tau-1) f (\tau) \right)\,, \\
A_W (\tau) &=& - \frac{1}{\tau^2} \left( 2 \tau^2 + 3 \tau + 3 (2 \tau - 1) f (\tau) \right)\,, \\
A_S (\tau) &=& - \frac{1}{\tau^2} \left( \tau - f (\tau) \right)\,;
\eeq
where
\beq
f (\tau) = \left\{ \begin{array}{lc}
\mbox{arcsin}^2 \sqrt{\tau}  & \tau \leq 1 \\
- \frac{1}{4} \left[ \log \frac{1+\sqrt{1-\tau^{-1}}}{1-\sqrt{1-\tau^{-1}}} - i \pi \right]^2 & \tau > 1 \end{array} \right.\,.
\eeq
For our study we are particularly interested in the limit of such functions for large mass of the particle in the loop with respect to the Higgs mass, $\tau \ll 1$:
\beq
A_F (0) = \frac{4}{3}\,, \quad
A_W (0) = - 7 \,, \quad
A_S (0) = \frac{1}{3}\,. 
\eeq
Note that the particle in the loop does not decouple for large mass because the (SM) coupling to the Higgs is also proportional to the mass of the particle.
As we are interested in Higgs masses below the $W$ threshold and above the LEP limit (where the $\gamma \gamma$ signal is non negligible), the light Higgs approximation is useful for the top and the new physics.
For the $W$, this approximation is not valid, and the function $A_W (\tau_W)$ ranges from $-8$ for $m_H = 115$ GeV to $-9.7$ for $m_H = 150$ GeV.

However, the mass of new particles in most models is not proportional to the Higgs VEV, but receives only a small correction from the electroweak symmetry breaking.
Therefore, the amplitude for new physics is given by the same formulae as above up to a factor taking into account the different coupling to the Higgs (which is in general not proportional to the mass).  
The coupling to the Higgs for a fermion (boson) can be written in general as~\footnote{Those formulae are valid for a SM Higgs sector. When the Higgs sector is extended, and for scalars which do mix with the Higgs doublet, more general formulae apply: see appendix~\ref{app:general} for details.}
\beq
y_{h\bar f f} = \frac{\partial m_f (v)}{\partial v}\,, \qquad y_{h\bar W W} = \frac{\partial m_W^2 (v)}{\partial v}\,.
\eeq
Therefore, without loss of generality, we can write the $A$ function for the new physics contribution as
\beq
A_{NP} = \frac{v}{m_{NP}} \frac{\partial m_{NP}}{\partial v}\; A_{F,W,S}\,.
\label{generalcoupling}
\eeq
When the mass of the new physics is not proportional to the Higgs VEV, $A_{NP}$ will decouple for large masses.

The new physics can be parametrised by two independent parameters describing the contribution of the new particles to the two decay widths, however using the actual amplitude is not a convenient way of treating the new contributions. Here we propose to normalise the new contribution to the top one.
The main reason is that the top gives the main contribution to the amplitudes in the SM, and any new physics, which addresses the problem of the Higgs mass naturalness, will have a tight relation with the top.
Moreover, as it will soon be clear, those two parameters can give some intuitive information about what kind of new physics runs into the loop. The widths can be rewritten as
\beq
\Gamma_{\gamma \gamma} &=& \frac{G_F \alpha^2 m_H^3}{128 \sqrt{2} \pi^3} \left| A_W (\tau_W) + 3 \left( \frac{2}{3} \right)^2 A_t (\tau_t)\; [1+\kappa_{\gamma \gamma} ] + \dots \right|^2\,, \\
\Gamma_{g g} &=& \frac{G_F \alpha_s^2 m_H^3}{16 \sqrt{2} \pi^3} \left| \frac{1}{2} A_t (\tau_t)\; [1+\kappa_{gg}] + \dots \right|^2\,,
\eeq
where the dots stand for the negligible contribution of the light quarks and leptons, and the coefficients $\kappa$ can be written as:
\beq
\kappa_{\gamma \gamma} &=& \sum_{NP}\;  \frac{3}{4} N_{c,NP} Q_{NP}^2\;  \frac{v}{m_{NP}} \frac{\partial m_{NP}}{\partial v}\; \frac{A_{F,W,S} (m_{NP})}{A_t}\,,\\
\label{kappagamma}
\kappa_{g g} &=& \sum_{NP}\;  2 C (r_{NP})\;  \frac{v}{m_{NP}} \frac{\partial m_{NP}}{\partial v}\; \frac{A_{F,W,S} (m_{NP})}{A_t}\,,
\label{kappaglu}
\eeq
where the ratio of $A$ functions depends on the spin and masses of the new particles (and top).
In the light Higgs approximation, however, the ratio only depends on the spin of the new particle:
\beq
 \frac{A_{NP}}{A_t}= \left\{ \begin{array}{l}
1 \quad \mbox{for fermions} \\
- \frac{21}{4} \quad \mbox{for vectors}\\
\frac{1}{4} \quad \mbox{for scalars}
\end{array} \right.
\eeq
An interesting feature of this parameterisation is that a particle with the same quantum numbers of the top will give $\kappa_{\gamma \gamma} = \kappa_{g g}$, and a single particle will give a contribution to the two coefficients with the same sign. 
In this way, if the experimental data allow to point to a specific quadrant in the $\kappagamma$--$\kappaglu$ parameter space, we can have a hint of the underlying new physics model. 
This will be illustrated in various examples in the following sections.
Note also that positive $\kappa$'s enhance the top contribution, therefore inducing an enhancement in the gluon channel but a suppression in the photon one, where there is a numerical cancellation between the dominant $W$ contribution and the top one.

The presence of new physics often modifies the tree level relation between the mass of the SM particles and the Higgs VEV.
This modification of the SM contribution can also be cast in the $\kappa$ parameters.
For the top it will read:
\beq
\kappa_{\gamma \gamma} (top) = \kappa_{gg} (top) =  \frac{v}{m_{t}} \frac{\partial m_{t}}{\partial v} -1\,.
\eeq
For the $W$:
\beq
\kappa_{\gamma \gamma} (W) &=& \frac{3}{4}\; \left( \frac{v}{m_{W}} \frac{\partial m_{W}}{\partial v} -1 \right)\; \frac{A_W (\tau_W)}{A_F (\tau_{top})}\,,\\
\kappa_{g g} (W) &=& 0\,.
\eeq

Note that the modification of the SM couplings will also affect the other production channels, and the branching ratio in heavy gauge bosons.
Those effects will however have a minor impact on our analysis, and their inclusion will be necessary in a later model-dependent analysis, after (and if) a model is preferred by data

\subsection{Observables at the LHC and Linear Colliders}

The LHC will measure the inclusive $\gamma \gamma$ Higgs decays and the new physics will modify both the total production cross section and the branching fraction in photons.
For large masses, close to the $W$ threshold, the decay in two heavy gauge bosons (one is virtual) becomes relevant and will also yield a relatively early measurement.
At large luminosities, one may also measure the $\gamma \gamma$ decays in a specific production channel, for instance the vector boson fusion one that can be isolated using two forward jet tagging: in this case one may probe directly the branching ratios.

In the Higgs mass range of interest, between 115 and 150 GeV, the main production channel is gluon fusion with a SM cross section of $40-25$ pb, followed by vector boson fusion ($5-4$ pb) and by other channels ($WH$, $ZH$, $\bar t t H$) which sum up to $4-2$ pb.
Here we will assume that the new physics significantly contributes only to the loop in the gluon fusion channel, while the other cross sections are unaffected.
The total production cross section normalised with the SM one, that we denote as $\bar \sigma$, can be written as:
\beq
{\bar \sigma}(H) = \left(\frac{\sigma_{gg}^{NP}+\sigma_{VBF}^{SM}+\sigma_{VH,\bar{t}tH}^{SM}}{\sigma_{gg}^{SM}+\sigma_{VBF}^{SM}+
\sigma_{VH,\bar{t}tH}^{SM}}\right)\simeq \left(\frac{(1+\kappaglu)^2 \sigma_{gg}^{SM}+\sigma_{VBF}^{SM}+\sigma_{VH,\bar{t}tH}^{SM}}{\sigma_{gg}^{SM}+\sigma_{VBF}^{SM}+\sigma_{VH,\bar{t}tH}^{SM}}\right) \label{sigmagg}\; .
\eeq

In the SM the Higgs branching fraction in photons amounts to $2 \cdot 10^{-3}$.
In presence of new physics, the branching fraction will also be sensitive to the gluon loop via the total width, as the gluon channel is significant: it amounts to 7\% of the total for $m_H = 115$ GeV, decreasing to 3\% for $m_H = 150$ GeV.
Also in this case, we define a branching ratio normalised to the SM value, $\overline {BR}$
\begin{multline}
{\overline {BR}}(H \rightarrow \gamma\gamma)= \frac{\Gamma_{\gamma \gamma}^{NP}}{\Gamma_{\gamma \gamma}^{SM}}\ 
\frac{\Gamma_{\rm tot}^{SM}}{\Gamma_{gg}^{NP}+\Gamma_{\gamma \gamma}^{NP}+\Gamma_{\rm others}^{SM}}  \\
\simeq \left( 1 + \frac{\kappagamma}{\frac{9}{16} A_W (\tau_W) + 1} \right)^2 \ \frac{\Gamma_{\rm tot}^{SM}}{(1+\kappaglu)^2\Gamma_{gg}^{SM}+(\Gamma_{\rm tot}^{SM} - \Gamma_{gg}^{SM})}\,.\label{brgg}
\end{multline}
The branching ratio in heavy vectors will depend on $\kappaglu$ via the total width of the Higgs, therefore the normalised $\overline {BR}$ is
\beq
{\overline {BR}}(H \rightarrow V V^*) =
\frac{\Gamma_{\rm tot}^{SM}}{\Gamma_{gg}^{NP}+\Gamma_{\gamma \gamma}^{NP}+\Gamma_{\rm others}^{SM}}  \simeq  \ \frac{\Gamma_{\rm tot}^{SM}}{(1+\kappaglu)^2\Gamma_{gg}^{SM}+(\Gamma_{\rm tot}^{SM} - \Gamma_{gg}^{SM})}\,.\label{brVV}
\eeq

For completeness, the normalised gluon branching fraction can be written as
\beq
{\overline {BR}}(H \rightarrow gg) = \frac{\Gamma_{gg}^{NP}}{\Gamma_{gg}^{SM}}\ 
\frac{\Gamma_{\rm tot}^{SM}}{\Gamma_{gg}^{NP}+\Gamma_{\gamma \gamma}^{NP}+\Gamma_{\rm others}^{SM}}\simeq \frac{( 1 + \kappaglu)^2  \Gamma_{\rm tot}^{SM}}{(1+\kappaglu)^2\Gamma_{gg}^{SM}+(\Gamma_{\rm tot}^{SM} - \Gamma_{gg}^{SM})}\,.\label{brglu}
\eeq
The branching ratios will be measured with an accuracy of few \% at a TeV $e^+e^-$ Linear Collider.

\section{Survey of models of New Physics in 4 dimensions}
\label{sec:models4}
\setcounter{equation}{0}

In this section we will summarise the values of the two parameters $\kappagamma$ and $\kappaglu$ in a variety of models of new physics.
It is not intended to be a complete survey, but rather a collection of examples of the usefulness of our proposed parametrisation, and of the impact of new physics on the Higgs search.
Here, we will briefly discuss a fourth generation, supersymmetry, Little Higgs models, a scalar colour octet and the Lee-Wick SM.
As the new particles and mass scales are often heavier than the top, we will use the light Higgs approximation to derive some simple analytical formulae.

\subsection{A 4$^{\rm th}$ generation}
\label{sec:4gen}

As for SM fermions, the masses of a chiral fourth generation are proportional to the Higgs VEV, and they cannot be arbitrarily large due to the perturbativity of the Yukawa couplings, naively $m_4 < 4 \pi v \sim 2$ TeV.
It has been shown that the impact of a relatively light 4$^{\rm th}$ generation on the electroweak precision tests can be minimised if the spectrum follows a specific pattern~\cite{4genKribs}: in particular if the splitting between the up and down type quarks is about 50 GeV (and similarly for the leptons).
For masses of a few hundred GeV, this is not a severe fine tuning.
Finally, let us remind that direct bounds on such new particles are of the order of 190 GeV (for a fourth generation bottom type quark in $p\bar p$ collisions \cite{pdg}) and 100 GeV for a charged lepton.

In the light Higgs approximation, the mass dependence disappears: $\kappaglu$ simply counts the number of new colour triplet quarks, $\kappaglu =2$, while $\kappagamma$ depends on the charges
\beq
\kappagamma = \frac{3}{4} \left[ 3 \left( \frac{2}{3} \right)^2 +  3 \left(- \frac{1}{3} \right)^2 + 1 \right] = 2\,.
\eeq
Due to an accident in the charges, therefore, a complete extra generation contributes like two tops.
Another accident is that the width in photons is largely suppressed, while the gluon one is enhanced by almost the same amount: overall, the inclusive $\gamma \gamma$ signal will be similar to the SM one~\cite{4genKribs} (for a light Higgs).

\subsection{Supersymmetry}
\label{sec:SUSY}

\begin{figure}[!t]
\begin{center}
\includegraphics[width=1\textwidth]{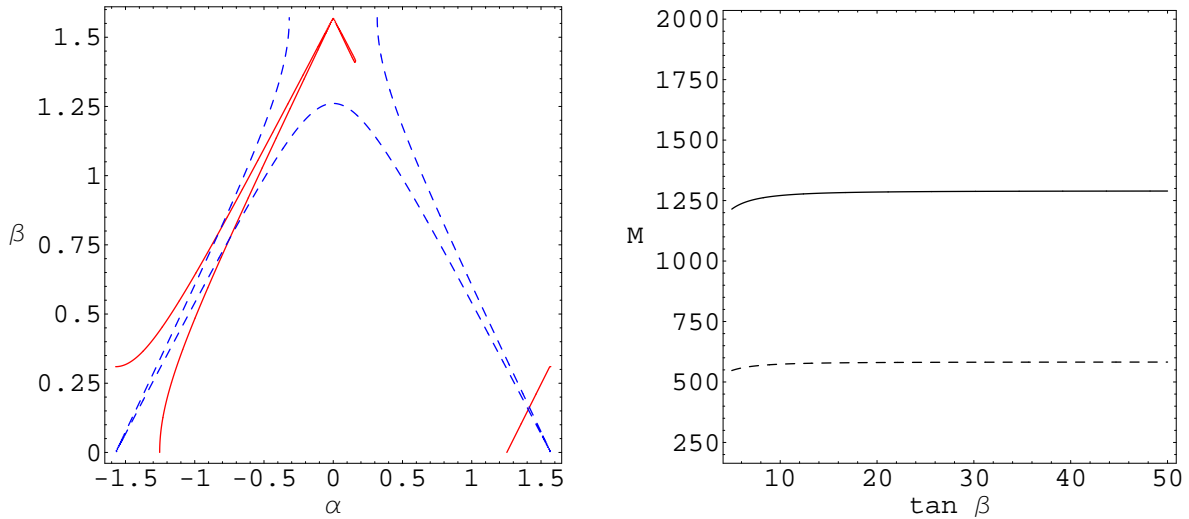}
\end{center}
\caption{\footnotesize Left panel: in solid red the region where the couplings of the $W$ and bottom are both within 5\% from the SM values (up to the sign), in dashed blue the same for the top couplings. Right panel: lower bound on the heavy higgs masses as a function of $\tan \beta$ requiring deviations of 1\% (solid) or 5\% (dashed).}
\label{fig:alphabeta}
\end{figure}

The supersymmetric contributions to the $h \to \gamma\gamma$ and $h \to g g$ amplitudes are well studied in supersymmetric extensions of the standard model (see for example \cite{Carena:2002qg} for few sample benchmark scenarios). 
Here we will focus on the common scenario where the heavier Higgses are above the $WW$ threshold, so that the $\gamma \gamma$ decay mode is only relevant for the light Higgs $h$.
However, the parametrisation we propose in this paper cannot be used in general for supersymmetric models.
In fact, due to the presence of two Higgses which develop a VEV, the tree level couplings of the SM particles to the Higgs are modified at order $\mathcal{O} (1)$ compared to the SM case.
If we define $\tan \beta = v_u/v_d$ the ratio of the two VEVs, and $\alpha$ the mixing angle in the neutral Higgs sector~\cite{susyprimer}, the couplings of $W$, top (up-type fermions) and bottom (down-type fermions) compared to the SM values are corrected by the following factors:
\beq
\frac{g_{W^+W^-h}}{g_{SM}} = \sin (\beta - \alpha)\,, \qquad \frac{g_{\bar{t}th}}{g_{SM}} = \frac{\cos \alpha}{\sin \beta}\,, \qquad \frac{g_{\bar{b}bh}}{g_{SM}} =- \frac{\sin \alpha}{\cos \beta}\,.
\eeq
Those corrections can be large, even for heavy susy masses. In the large $\tan \beta$ case, which is preferred by the top Yukawa perturbativity and experimental constraints, the bottom (and tau) Yukawas are enhanced by a large factor $\sim \tan \beta$: the Higgs width increases and the branching ratio in photons can be easily suppressed by orders of magnitudes, making this channel unobservable.
In order to keep the $\gamma \gamma$ channel alive, one needs to compensate the large $\tan \beta$ with a small mixing angle in the Higgs sector: $\alpha \sim \pm ( \pi/2-\beta)$.
In order to safely use our formalism, we need to make sure that the corrections to the bottom Yukawa (and couplings to the $W$) are negligible.
In the left panel of Figure~\ref{fig:alphabeta} we plotted the region in the $\alpha$--$\beta$ parameter space where both the $W$ and bottom couplings deviate by less than 5\% from the SM value (up to the overall sign).
We also superimposed the region where corrections to the top Yukawa are smaller than 5\%.
There is a tiny region where a fine tuning between the two angles allow for our formalism to be used.
Note that a larger mixing angle in the Higgs sector will soon enhance the bottom Yukawa and kill the $\gamma \gamma$ signal, so the region is not a negligible part of the parameter space where the signal is observable.
On the right panel we used a three level relation between the masses in the Higgs sector and $\alpha$
\beq
\frac{\tan 2 \alpha}{\tan 2 \beta} = \frac{m_A^2 + m_Z^2}{m_A^2 - m_Z^2}\,,
\eeq
where $m_A$ is the mass of the pseudoscalar (which also sets the mass scale of the other heavy Higgses), to set a lower bound on the heavy Higgs masses.
Therefore, we expect that for masses above $1$ TeV, the corrections to the bottom (and tau) Yukawa can be safely neglected.
In this region, corrections to the $W$ and top couplings are small too.

For the purpose of illustrating our parameterisation, we will focus on some approximate expressions that arise in a simple scenario: the MSSM golden region~\cite{Perelstein:2007nx}.
This scenario is motivated by naturalness in the Higgs mass, minimal fine tuning and precision tests.
The main features are large soft masses for the gauginos and for the light generations, and large mixing in the stop sector induced by a large soft trilinear term.
A general analysis of the $\gamma \gamma$ channel can be found in Ref.~\cite{low}.
As a numerical example we will consider a variation of the benchmark point in Ref.~\cite{Perelstein:2007nx}: here, $\tan \beta = 10$ and all the soft masses except the stop and Higgs ones are at 1 TeV, $\mu = 250$ GeV and the soft trilinear term for the stops $A_t$ is at 1 TeV to induce a large mixing in the stop sector and reduce the fine tuning in the Higgs potential.
In this benchmark point, the light Higgs is at $129$ GeV. 
Charginos and neutralinos (mostly higgsinos) are at 250 GeV (set by $\mu$), while the stops are at 400 and 700 GeV.
All other masses are above a TeV, and we will neglect the contribution of those sparticles.
The only difference is that, in order to avoid the bottom Yukawa problem, we will push the heavy Higgses above 1 TeV: to do that it is enough to increase the $H_d$ soft mass above the TeV scale.
This will not introduce a severe fine tuning, as the contribution of this mass to the Higgs VEV is also suppressed by the large $\tan \beta$~\cite{susyprimer}. 
In this scenario, only the stops contribute to $h \to gg$ and $h \to \gamma \gamma$. 
We neglect the contribution of charginos because they are mostly higgsinos: the coupling to the Higgs is suppressed by the large gaugino masses.
For the stops, assuming that the soft masses for left and right handed scalars are equal ($\sim 550$ GeV at the benchmark point), the contribution to the $\kappa$ parameters can be expressed as
\beq
\kappaglu ({\mathrm stops}) = \kappagamma ({\mathrm stops}) \simeq \frac{m_2^2 + m_1^2}{4 m_1^2 m_2^2} m_t^2 - \frac{(m_2^2 - m_1^2)^2}{16 m_1^2 m_2^2} \sim -0.02\,,
\eeq
where $m_{1,2}$ are the masses of the two eigenstates, and the second term is proportional to the soft trilinear term: $m_2^2 - m_1^2 \simeq 2 |A_t| m_t$.

Those formulae are presented here for illustration purpose only, and we will use exact one loop expressions for the numerical analysis, including the contribution of charginos.
A more general analysis of the region in the MSSM parameter space is beyond the scope of this paper, and it is postponed to a following publication.

\subsection{Little Higgs models}

In Little Higgs models~\cite{LH}, the gauge symmetries of the SM are a subgroup of a larger global symmetry.
The breaking of such symmetry at a higher scale $f$ produces light pseudo-Goldstone bosons, which we want to identify with the Higgs boson.
The symmetry structure removes the divergences from the Higgs mass at one loop: the reason being that one loop is not sensitive to the explicit breaking of such global symmetry (while higher loops are).
This is however enough to solve the little hierarchy problem, because the scale of new physics required beyond the Little Higgs mechanism is pushed above 10 TeV.
In general new gauge bosons are introduced in order to eat up unwanted Goldstone bosons, and they also generate the loops that cancel the divergences from the SM gauge bosons. 
Similarly, new fermionic states, cousins of the top, required by the global symmetry, will cancel the divergences of the top loop.

We will first derive some very general formulae, and then apply them to explicit examples.
In models with only one extra massive gauge boson, $W$', the cancellation works thanks to the different sign between the couplings of the $W$ and $W$'~\cite{LHreview}:
\beq
g_{hW'W'} = - g_{hWW}\,.
\eeq
This is a consequence of the fact that $m_W^2 + m_{W'}^2$ does not depend on the Higgs VEV, but it is fixed by the scale $f$ at which the global symmetry is broken.
The coupling of the $W$ with the Higgs is also modified
\beq
g_{hWW} = g_{hWW}^{SM}\; (1-\delta_W) = \frac{2 m_W^2}{v}\; (1-\delta_W) \,,
\eeq
where $\delta_W$ contains corrections in $v/f$.
After recalling that $g_{hWW} = \frac{\partial m_W^2}{\partial v}$, the contribution of the $W$ and $W$' to the $\kappa$ parameters is (in the light Higgs approximation) :
\beq
\kappagamma (W') &=& \frac{63}{16} \left( \frac{m_W}{m_{W'}} \right)^2 (1-\delta_W)\,,\\
\kappagamma (W) &=& - \frac{9}{16} A_W (\tau_W) \delta_W\,;
\eeq
while $\kappaglu=0$.

The precise value of $\delta_W$ depends on the symmetry structure of the model: in the Simplest Little Higgs (SLH) model~\cite{simplest}, which is based on an SU(3) gauge symmetry,
\beq
m_W & = & g f \sin \frac{v}{2 f}\,, \\
m_{W'} &=& g f \cos \frac{v}{2 f}\,;
\eeq
so that 
\beq
\delta_W = 1- \frac{m_{W'}}{m_{W}} \mbox{arctan} \, \frac{m_W}{m_{W'}} \simeq \frac{1}{3} \left( \frac{m_W}{m_{W'}} \right)^2 + \dots
\eeq
At leading order in $m_W/m_{W'} \sim v/f$:
\beq
\kappagamma \simeq \frac{9}{16} \left( 7 - \frac{1}{3}  A_W (\tau_W) \right)\;\left(  \frac{m_W}{m_{W'}} \right)^2 \sim (6.2 \div 6.7)\, \cdot\, \left(  \frac{m_W}{m_{W'}} \right)^2
\eeq
where we have varied the Higgs mass between 115 and 150 GeV.

The top sector is more complicated because doubling of fields is usually required in order to generate a realistic spectrum for the light states.
For instance, the simplest way to introduce the top is to embed the SM left-handed doublet in a triplet of SU(3) that couples via the two Higgses to two right-handed singlets.
In this case we need to double the right-handed tops in order to give mass both to the top and to its heavy partner $T$.
The symmetry structure of the model implies that $m_t^2 + m_T^2$ does not depend on the SM Higgs VEV, therefore the following relation holds between the couplings to the Higgs ($g_{hff} = \frac{\partial m_f}{\partial v}$):
\beq
g_{hTT} = - \frac{m_t}{m_T} g_{htt}\,.
\eeq
As in the gauge sector, the coupling of the SM top also receives deviations from the usual Yukawa coupling, which we can parametrise as
\beq
g_{htt} = \frac{m_t}{v} (1-\delta_t)\,.
\eeq
In terms of this parameterisation, the contribution to $\kappagamma$ and $\kappaglu$ of the top and $T$ are, in the light Higgs approximation:
\beq
\kappagamma (top) = \kappaglu (top) &=& - \delta_t\,,\\
\kappagamma (T) = \kappaglu (T) &=& - \frac{m_t^2}{m_T^2} (1- \delta_t)\,.
\eeq

In the SLH model, the masses can be written as
\beq
m_{t,T}^2 &=& \lambda_T^2 f^2 \left( 1 \mp \sqrt{1-\frac{\lambda_t^2}{\lambda_T^2} \sin^2 \frac{v}{f} } \right)\,,
\eeq
where $\lambda_{t,T}$ are related to the Yukawa couplings to the two Higgses $\lambda_{1,2}$:
\beq
\lambda_T = \sqrt{\frac{\lambda_1^2 + \lambda_2^2}{2}}\,, \qquad \lambda_t = \frac{\lambda_1 \lambda_2}{\lambda_T}\,.
\eeq
Therefore
\beq
\delta_t = 1-\frac{m_T^2}{m_T^2 - m_t^2} \frac{m_{W'}^2 - m_W^2}{m_{W'} m_W} \mbox{arctan} \frac{m_W}{m_{W'}} \simeq - \frac{m_t^2}{m_T^2} + \frac{4}{3} \frac{m_{W}^2}{m_{W'}^2} + \dots
\eeq
and, at leading order,
\beq
\kappagamma (top+T) = \kappaglu (top+T) \simeq - \frac{4}{3}  \frac{m_{W}^2}{m_{W'}^2} + \dots
\eeq
Note that at leading order the result is independent on the heavy top mass, but only depends on the heavy gauge boson $W'$.

The total contribution is therefore:
\beq
\kappaglu (SLH) &\simeq & - \frac{4}{3} \frac{m_W^2}{m_{W'}^2} \sim -0.002 \cdot \left( \frac{2 \mbox{TeV}}{m_{W'}} \right)^2\,, \\
\kappagamma (SLH) &\simeq & \left( \frac{47}{12} - \frac{3}{16} \left( 7+A_W \right) \right)  \frac{m_W^2}{m_{W'}^2} \sim 0.006 \cdot \left( \frac{2 \mbox{TeV}}{m_{W'}} \right)^2 \,;
\eeq
the Higgs mass dependence in $A_W$ is very mild due to the small coefficient.
In the numerical values we have chosen a $W'$ mass of 2 TeV, which is roughly the one required by electroweak precision measurements~\cite{SLHprecision}. Note however that the implementation of a T parity~\cite{Tparity} would reduce the bound by almost an order of magnitude.

Another simple model using the Little Higgs mechanism was proposed in Ref.~\cite{littlest} and dubbed Littlest Higgs.
Here a global SU(5) is spontaneously broken down to SO(5), and a subgroup SU(2)$^2 \times$ U(1)$^2$ is gauged.
The mechanism acts thanks to the presence of two copies of the SM gauge group, which are broken to the diagonal by the spontaneous breaking of SU(5).
The Higgs again is a pseudo-Goldstone boson of the global symmetry breaking.
The model, together with a heavy $W$ ($W_H$) and top ($T$) also contains a heavy charged scalar $\Phi$ from a triplet of SU(2) that develops a VEV ($v'$).
The model therefore contains more parameters that the SLH, and its contribution to $H\to \gamma \gamma$ and $H\to gg$ has been computed in Ref.~\cite{LLHgamma}.
Here we will simply translate those results in our notation: the contribution of the $W$ and heavy gauge states is (expressed in terms of the masses at leading order in the Higgs VEV $v$):
\beq
\kappagamma (W) &\simeq& \frac{9}{16} \left( \frac{m_W^2}{m_{W_H}^2} - \frac{5-x^2}{8}  \frac{v^2}{f^2} \right) A_W (\tau_W)\,, \\
\kappagamma (W_H) &\simeq& \frac{63}{16} \frac{m_W^2}{m_{W_H}^2}\,, \\
\kappagamma (\Phi) &\simeq& - \frac{4-3 x^2}{64}  \frac{v^2}{f^2}\,; \\
\kappaglu (W, W_H, \Phi) &=& 0\,.
\eeq
Here $x$ is proportional to the ratio between the triplet and doublet VEVs ($0 \leq x < 1$).
Note that in the limit where the Higgs is much lighter than the $W$-threshold, the contributions proportional to the $W$ mass tend to cancel, while the ones proportional to $v^2$ do not.
The top and heavy top contributions are
\beq
\kappagamma (top) = \kappaglu (top) &\simeq&  \frac{m_t^2}{m_T^2} - \frac{7-4x+x^2}{8} \frac{v^2}{f^2}\,, \\
\kappagamma (T) = \kappaglu (T) &\simeq& - \frac{m_t^2}{m_T^2}\,.
\eeq
Note that as in the SLH, the contribution proportional to the top mass cancels out.
Therefore, the main corrections in this model are proportional to the Higgs VEV and suppressed by the global symmetry breaking $f$:
\beq
\kappaglu (LH) &\simeq& - \frac{7-4 x+x^2}{8} \frac{v^2}{f^2}\,, \\
\kappagamma (LH) &\simeq& \left( \frac{195}{64} +x-\frac{73}{64} x^2 \right)  \frac{1}{2}  \frac{v^2}{f^2} +  \nonumber \\
 & & +\frac{9}{16} \left( 7+ A_W \right) \left( \frac{m_W^2}{m_{W_H}^2} - \frac{5-x^2}{8}  \frac{v^2}{f^2} \right)\,.
\eeq
Note that the second term in $\kappagamma$ , which depends on the $W_H$ mass, is negligible due to a small coefficient, therefore the result only depends on $f$ and the triplet VEV $x$.

The bound from precision measurements on the scale $f$ is around 5 TeV~\cite{SLHprecision}, which correspond roughly to masses of order 2 TeV.
To give a numerical example, if $x$ is negligible
\beq
\kappaglu (LH) &\simeq& - \frac{7}{8} \frac{v^2}{f^2} \sim -0.002 \cdot \left(\frac{5 TeV}{f}\right)^2\,, \\
\kappagamma (LH) &\simeq& \frac{195}{128} \frac{v^2}{f^2} \sim 0.0036 \cdot \left(\frac{5 TeV}{f}\right)^2\,.
\eeq
Not that when a T parity is implemented on this model~\cite{Tparity}, the bound on $f$ is lowered to $500$ GeV~\cite{LLHprecision}, therefore the contribution to the $\kappa$ parameters is 100 times bigger.

\subsection{Extended scalar sector: colour octet} \label{sec:coloctet}

The scalar sector is experimentally the least tested part of the Standard Model and may be more complicated than the minimal content of the SM.
It has been shown~\cite{Manohar:2006ga} that in order to avoid tree level flavour changing neutral currents, the extra scalar should be either a copy of the SM one (leading to the two Higgs model) or a colour octet with the same weak quantum numbers as the SM Higgs.
Here we will focus on the latter possibility~\cite{Manohar:2006ga}.
The most general potential contains 3 terms that are bilinear in both the Higgs $H$ and the colour octet $S$:
\beq
\mathcal{L} = \lambda_1 H^{\dagger i} H_i S^{\dagger j} S_j + \lambda_2 H^{\dagger i} H_j S^{\dagger j} S_i + \left[ \lambda_3 H^{\dagger i} H^{\dagger j} S_i S_j + h.c. \right] + \dots
\eeq
where $i$ and $j$ are SU(2) indices and we have left implicit the colour contractions.
Note that imposing custodial symmetry would require $\lambda_2 = 2 \lambda_3$.
After the Higgs develops a VEV $\langle H \rangle = v/\sqrt{2}$, the spectrum contains one charged and two neutral scalar octets with masses
\beq
m^2_{S^{\pm}} &=& m_S^2 + \lambda_1 \frac{v^2}{4} = m_S^2\, (1+X_1)\,, \\
m^2_{S^0_{1,2}} &=& m_S^2 + (\lambda_1 + \lambda_2 \pm 2 \lambda_3)  \frac{v^2}{4}  = m_S^2\, (1+X_1 +X_2\pm 2 X_3)\,;
\eeq
where $X_i = \lambda_i v^2/4$.
At loop level, the octet will  contribute to the electroweak precision tests~\cite{Manohar:2006ga}: the corrections can be encoded in the $S$ and $\rho$ parameters, and for small $v \ll m_S$:
\beq
S &\simeq& \frac{2}{3 \pi} X_2\,, \\
\Delta \rho &\simeq& \frac{\sin^2 \theta_W\, m_W^2}{96 \alpha \pi^3\, m_S^2} (\lambda_2^2 - 4 \lambda_3^3 )\,.
\eeq
The corrections to the $\rho$ parameter can be minimised by imposing (approximate) custodial symmetry, while $S$ will give a direct constraint on $X_2$.
Note that $X_1$ is not strongly constrained.

Using the formalism developed in the previous section we can compute the contribution of the scalar octet to the $\kappa$ parameters (for $v\ll m_S$):
\beq
\kappagamma (S) &\simeq& \frac{3}{2} \frac{\lambda_1 v^2}{4 m_{S^\pm}^2} \sim \frac{3}{2}     X_1 \,, \\
\kappaglu (S) &\simeq & \frac{C (8)}{2} \left( \frac{\lambda_1 v^2}{4 m_{S^\pm}^2} + \frac{1}{2} \frac{(\lambda_1+\lambda_2 + 2 \lambda_3) v^2}{4 m_{S^0_1}^2} + \frac{1}{2} \frac{(\lambda_1+\lambda_2 - 2 \lambda_3) v^2}{4 m_{S^0_2}^2}\right) \nonumber \\
 &\sim& \frac{3}{2} (2 X_1 + X_2)\,;
\eeq
where $C (8) = 3$.
As a numerical example, we will use $\lambda_1=4$, $\lambda_2=1$ and $m_S=750$ GeV.
In this case, $X_1 \sim 1/9$ and $X_2 \sim 1/36$, therefore $\kappagamma \sim 0.17$ and $\kappaglu \sim 0.37$.

\subsection{Lee-Wick Standard Model}

Lee and Wick (LW) proposed a modification of the particle propagators in QED by means of higher derivative terms in order to improve the ultraviolet convergence of the theory and make loop corrections finite.
This modification of the propagator can also be parametrised by the presence of a new degree of freedom with large mass and negative kinetic term, so that the corrected propagator looks like a Pauli-Villars regularised one, where the Pauli-Villars cutoff scale is replaced by the mass scale of such new degree of freedom.
This idea has recently been extended to the full SM~\cite{Grinstein:2007mp}: in this case the loops are not finite, however the softening of the divergences is enough to address the hierarchy problem in the Higgs mass.

Notwithstanding the theoretical issues arisen by this formulation, the contribution of the LW degrees of freedom to the $H \to g g$ and $H \to \gamma \gamma$ amplitudes has been computed~\cite{Krauss:2007bz}: here we will sketch the calculation, making use of the general formulae given in Section~\ref{sec:definitions}, and give some simple results in the large LW mass approximation.

In this model, to each SM particle, a new LW degree of freedom is associated (2 for each chiral fermion).
For more details of the construction we refer the reader to the Refs.~\cite{Grinstein:2007mp,Krauss:2007bz}.
The Higgs VEV will generate a mixing between the standard and LW particles, which has been studied in detail in~\cite{Krauss:2007bz}: in the following we will review just the results needed to complete our calculation.

In the Higgs sector, the SM Higgs $h$ and the LW scalar $\tilde h$ mix via the Higgs VEV: the mixing can be described by a symplectic rotation
\beq
\left( \begin{array}{cc}
\cosh \theta & \sinh \theta \\
\sinh \theta & \cosh \theta
\end{array} \right)\,, \qquad \mbox{with} \quad \tanh 2 \theta = - 2 \frac{m_h^2 \tilde{m}_{ h}^2}{m_h^4 + \tilde{m}_{h}^4}\,.
\eeq
A very similar mixing takes place in the gauge sector, between the $W$ and the LW $\tilde W$.
The two mass eigenstates are
\beq
m_W^2 & = & \frac{1}{2} \left( M_2^2 - \sqrt{M_2^4 - g^2 v^2 M_2^2} \right)\,, \\
\tilde{m}_W^2 & = & \frac{1}{2} \left( M_2^2 + \sqrt{M_2^4 - g^2 v^2 M_2^2} \right)\,;
\eeq
where $M_2$ is the mass of the LW partner of the SU(2) gauge bosons.
Note that there is no trilinear coupling between the $W$ and the LW Higgs $\tilde h$, therefore (using the formulae in Appendix~\ref{app:general}):
\beq
\frac{v}{m_W} \frac{\partial m_W}{\partial v} = \cosh \theta\, \frac{g^2 v^2}{4 m_W^2} \frac{M_2^2}{\sqrt{M_2^4 - g^2 v^2 M_2^2}} = \frac{\tilde{m}_h^2}{\sqrt{m_h^4+ \tilde{m}_h^4}} \frac{\tilde{m}_W^2}{\tilde{m}_W^2 - m_W^2} \simeq 1+\frac{m_W^2}{\tilde{m}_W^2}\,.
\eeq
For the $\tilde W$, the coupling to the Higgs is given by $- \frac{\partial \tilde{m}_W^2}{\partial v}$: the minus sign comes from the negative sign of the kinetic term.
This can be proved by an explicit calculation, and it is true for all the LW fields (for fermions, the coupling to the Higgs is $ - \frac{\partial \tilde{m}_f}{\partial v}$).
However, another minus sign comes from the loops: compared to the SM ones, propagators and couplings to the gauge bosons (photons and gluons) have a minus sign from the negative kinetic term of the LW fields.
All in all, a minus sign form the loop compensates the minus sign in the Higgs coupling and we can safely use the formulae in Section~\ref{sec:definitions}:
\beq
\frac{v}{\tilde{m}_W} \frac{\partial \tilde{m}_W}{\partial v} = - \cosh \theta\, \frac{g^2 v^2}{4 \tilde{m}_W^2} \frac{M_2^2}{\sqrt{M_2^4 - g^2 v^2 M_2^2}} = - \frac{\tilde{m}_h^2}{\sqrt{m_h^4+ \tilde{m}_h^4}} \frac{m_W^2}{\tilde{m}_W^2 - m_W^2}\simeq -\frac{m_W^2}{\tilde{m}_W^2}\,.
\eeq
Putting the two results together, and expanding for $m \ll \tilde{m}$:
\beq
\kappagamma (W + \tilde W) \simeq \frac{9}{16} \frac{m_W^2}{\tilde{m}_W^2} \left( 7+A_W (\tau_W) \right)\,.
\eeq
This result is numerically small, and the contribution of the $W$ and its LW partner tend to cancel each other in the light Higgs limit.
This result contradicts the findings of Ref.~\cite{Krauss:2007bz}~\footnote{The authors of Ref.~\cite{Krauss:2007bz} find that the contribution of the $W$ is proportional to the SM amplitude by a factor $s_{(A+\tilde{A})^2} = \frac{\tilde{m}_W^2 + m_W^2}{\tilde{m}_W^2 - m_W^2}$. In fact, the coupling of the $W$ with the Higgs can be written as $\cosh \theta\,s_{(A+\tilde{A})^2}\, \frac{g^2 v}{2}$; however the authors do not take into account that $\frac{g^2 v^2}{4 m_W^2} = \frac{m_W^2}{\tilde{m}_W^2 + m_W^2} \neq 1$.}. 

The spectrum also contains a LW charged scalar: in fact the LW Higgs does not develop a VEV and its charged component $\tilde{h}^+$ is not eaten up.
The mass of such scalar is simply given by the Higgs LW mass $\tilde{m}_{h^\pm}^2 = M_H^2$.
Nevertheless, as it happens in the SM with the Goldstone boson in the Higgs, the Lagrangian contains a coupling between $\tilde{h}^+$ and the Higgs field which can be calculated explicitly and enters the formulae in section~\ref{sec:definitions} as (we are using here the same notation as in Ref.s~\cite{Grinstein:2007mp,Krauss:2007bz})
\begin{multline}
\frac{v}{\tilde{m}_{h^\pm}} \frac{\partial \tilde{m}_{h^\pm}}{\partial v} \to -  \frac{v}{2 \tilde{m}_{h^\pm}^2} (\cosh \theta - \sinh \theta) \frac{ \lambda v}{2} = \\
 - \frac{1}{2} \frac{m_h^2 + \tilde{m}_h^2}{\sqrt{m_h^4 + \tilde{m}_h^4}} \frac{m_h^2 \tilde{m}_h^2}{\tilde{m}_{h^\pm}^2 (m_h^2 + \tilde{m}_h^2)} \simeq - \frac{1}{2} \frac{m_h^2}{\tilde{m}_{h^\pm}^2}\,,
\end{multline}
where an extra minus sign comes from the propagators in the loop.
Therefore
\beq
\kappagamma (\tilde{h}^\pm) \simeq - \frac{3}{32} \frac{m_h^2}{\tilde{m}_{h^\pm}^2}\,.
\eeq
This result is also different from the result in Ref.~\cite{Krauss:2007bz}, where the contribution of the charged LW Higgs vanishes at this order~\footnote{In Ref.~\cite{Krauss:2007bz}, the authors include the contribution of the charged LW Higgs using the amplitude of the SM Goldstone boson in the Feynman gauge rescaled by the ratio $\frac{m_W^2}{\tilde{m}_{h^\pm}^2}$. However, this naive estimation is incorrect, because, the coupling of a Goldstone boson (which is the same as $\tilde{h}^+$) is not proportional to its mass. See Appendix~\ref{app:GB} for more details.}.

The top sector is more complicated because for each chiral SM fermion one needs to add a massive Dirac fermion (with negative kinetic term).
The Yukawa couplings, however, have a simple form: in particular they have the same structure as the SM Yukawas, and they are functions of the field combination $H - \tilde H = 1/\sqrt{2} (v + h - \tilde{h} + \dots )$: the presence of a LW Higgs will only manifest itself in the fact that the couplings to the standard Higgs are proportional to $\cosh \theta - \sinh \theta$.
The spectrum can be calculated as a series for large LW top mass $M_t$ (assuming the same mass for the LW partners of the left- and right-handed tops):
\beq
m_t &=& M_t \epsilon \left( 1 + \epsilon^2 + \dots \right)\,,\\
\tilde{m}_{t1,2} &=& M_t \left( 1 \mp \frac{1}{2} \epsilon - \frac{3}{8} \epsilon^2 + \dots \right)\,;
\eeq
where $\epsilon = \frac{y_t v}{\sqrt{2} M_t}$.
The contribution of the top (and partners) is therefore:
\begin{multline}
\kappaglu ({\rm tops}) = \kappagamma ({\rm tops}) = \left( (\cosh \theta - \sinh \theta) \frac{\epsilon}{m_t} \frac{\partial m_t}{\partial \epsilon} -1 \right) + \\(\cosh \theta - \sinh \theta) \left( \frac{\epsilon}{\tilde{m}_{t1}} \frac{\partial \tilde{m}_{t1}}{\partial \epsilon} +\frac{\epsilon}{\tilde{m}_{t2}} \frac{\partial \tilde{m}_{t2}}{\partial \epsilon}\right)
 \simeq \frac{m_h^2}{\tilde{m}_h^2}\,.
\end{multline}
The $\epsilon$ dependence cancels out between the top and LW tops contributions, at the end the result only depends on the LW Higgs mass.

In total (at this order $\tilde{m}_{h^\pm} = \tilde{m}_h$):
\beq
\kappaglu & \simeq & \frac{m_h^2}{\tilde{m}_h^2} \sim 0.014\cdot \left( \frac{m_h}{120\, {\rm GeV}} \frac{1\, {\rm TeV}}{\tilde{m}_h} \right)^2\,, \\
\kappagamma & \simeq & \frac{29}{32} \frac{m_h^2}{\tilde{m}_h^2} + \frac{9}{16} \frac{m_W^2}{\tilde{m}_W^2} \left( 7+A_W (\tau_W) \right) \sim 0.013 \cdot \left( \frac{m_h}{120\, {\rm GeV}} \frac{1\, {\rm TeV}}{\tilde{m}_h} \right)^2\,.
\eeq
Those contributions are numerically much smaller than the results in Ref.~\cite{Krauss:2007bz}.

\section{Survey of models of New Physics in extra dimensions}
\label{sec:models5}
\setcounter{equation}{0}

In this section we will focus on models of new physics in one extra dimension, in particular on the different ways one can employ the Higgs mechanism in this context.
Most models can be divided in 3 main categories: bulk Higgs (BH), brane Higgs (bH) and Gauge Higgs (GH).
In the first case, the Higgs is just a 5D scalar field in the bulk, which picks up a VEV due to a potential, which may be localised on one brane.
In this class of models we find Universal Extra Dimensions~\cite{BHN,Cacciapaglia:2001nz,UED} and gaugephobic Higgs models~\cite{gaugephobic} in warped space, as an example.
In brane Higgs models, the Higgs is a 4 dimensional field localised on one brane or end-point of the compact space: the advantage of these models is that there is no tower of massive scalars and, if the brane where the Higgs is localised plays a special role like the TeV brane in warped space, the model may address the little hierarchy problem. A model of this kind was proposed by Randall-Sundrum~\cite{RShiggs}.
Finally, a new possibility allowed only in extra dimensional models is that the Higgs is part of a gauge group~\cite{hosotani}: in fact the 5$^{\rm th}$ component of a bulk gauge vector is a scalar from the 4D point of view.
The interactions and potential of such particle are however constrained by 5D Lorentz and gauge invariance: in particular, the Higgs potential (including its mass) is finite and insensitive to the physics at the cutoff.
The limit of this mechanism is that the model is only valid below an effective scale of few TeV (few Kaluza-Klein modes).
In this class we can find examples both in flat~\cite{GHUflat1,GHUflat2,MaruOkada} and warped space~\cite{GHUwarped,Djouadi,falkowski}.

Extra dimensional models are by nature non normalisable: from the 4D point of view, they are an effective description of the physics below a cutoff scale where some of the bulk interactions become strong.
Such scale lies typically above a few tens of Kaluza-Klein (KK) modes.
In general, if the symmetries allow so, we can add a tree level higher order operator which describes the coupling between the Higgs and the massless gauge bosons: in this way, the decay widths would be non-calculable.
Adding such operator is actually necessary in order to act as a counter-term to the divergences that will arise at loop level.
However, the loops we are interested in are effectively a box diagram if one considers a VEV insertion in the loop, therefore the one loop calculation turns out to be finite in all 5 dimension models.
The counter-term will only be required at higher loops, and we will take the finite one loop result as a good approximation.
In some cases, like in the Barbieri-Hall-Nomura model~\cite{BHN}, the operator is actually forbidden by an extra symmetry (supersymmetry in this case).
Models of Gauge Higgs are special: the Higgs interactions are constrained not only by gauge symmetry, but by 5D Lorentz invariance as well.
This is enough to forbid a tree level potential for the Higgs, and also tree level contributions to the decay widths.
Therefore, in this models the Higgs mass is really protected by symmetry and our calculation can be trusted as UV insensitive~\cite{Maru}.

We will present some general results on two different geometries: a flat extra dimension compactified on an interval (which is equivalent to an orbifold) and a warped extra dimension.

\subsection{Gauge bosons in a flat extra dimension}

In the flat case, the metric is an extended Minkowski, where the extra coordinate $y$ lies on an interval $[0, \pi L]$.
The notation is such that typically the mass of the first Kaluza-Klein state is $m_{KK} = 1/L$: this will be our reference mass scale in the following.
Note that this is the only mass scale introduced by the extra space structure.
This scale should be much larger than the $W$ mass due to direct and indirect constraints: the electroweak precision tests usually push it above $\sim 2$ TeV (see for example \cite{GHUflat1,GHU2precision}).
Is it possible to relax this bound by adding symmetries: as a typical number in this scenario we will use $m_{KK} \sim 500$ GeV.
This is the case, for example, in Universal Extra dimensions due to a Kaluza-Klein parity or the BHN model.

\subsubsection{Gauge Higgs}

One of the peculiarities of this models is the presence of a tower of charged vectors, $H_\mu^+$, associated with the charged component of the Higgs.
They will necessarily mix with the $W_\mu$ via the Higgs VEV.
Here we will focus on the simplest example, a $SU(3)$ gauge symmetry in the bulk, broken to $SU(2)\times U(1)$ at both endpoints.
The value of the Higgs VEV can be expressed in terms of a dimensionless parameter $\alpha$, which is indeed proportional to the field expectation value.
We postpone all details of the calculation of the spectrum and the precise definition of $\alpha$ in the Appendix~\ref{app:gaugehiggs}.
The spectrum for the $W$ and its KK tower is simply given, in terms of $\alpha$, as
\beq
m_n^2 = \frac{(n + \alpha)^2}{L^2}\,, \quad n = 0, \pm 1, \pm 2 \dots
\eeq
where $n=0$ corresponds to the $W$ mass:
\beq
m_W = \frac{\alpha}{L}\,.
\eeq
For the purpose of this section, this can be considered as the definition of $\alpha = m_W L = m_W/m_{KK}$: it is typically a small number because we want the mass of the first KK mode to be much larger than the $W$ mass in order to avoid direct and indirect bounds.
The $W$ mass is proportional to the Higgs VEV, so that its contribution to the loop is equal to the SM one.
The KK tower contribution to the $\kappa$'s is proportional to~\footnote{Note that, because $\alpha$ is proportional to the Higgs VEV, $\frac{\alpha}{m_n} \frac{\partial m_n}{\partial \alpha} = \frac{v}{m_n} \frac{\partial m_n}{\partial v}$.}
\beq
\sum_{n} \frac{\alpha}{m_n} \frac{\partial m_n}{\partial \alpha} =
\alpha \sum_{n=1}^{\infty} \frac{1}{n+\alpha} - \frac{1}{n-\alpha}
= \pi \alpha \cot \pi \alpha -1 = - \frac{\pi^2 \alpha^2}{3} + \mathcal{O} (\alpha^4)\,.
\eeq
We can use the definition of $\alpha$ to express the result in terms of the $W$ mass and the mass of the first KK mode $m_{KK} = 1/L$:
\beq
\kappa_{\gamma \gamma } (W_{KK}) &=& - \frac{63}{16} \left( \pi \frac{m_W}{m_{KK}} \cot \left( \pi \frac{m_W}{m_{KK}}\right) -1 \right) \nonumber\\
 &\simeq& \frac{63}{16} \frac{\pi^2}{3} \frac{m_W^2}{m_{KK}^2} \sim 0.021\cdot \left( \frac{2\, \mbox{TeV}}{m_{KK}} \right)^2\,;
\eeq
and $\kappa_{gg} = 0$.
Note that the contribution has an opposite sign compared to the $W$.

Models with a gauge group larger than $SU(3)$ may also contain gauge bosons with different boundary conditions on the two endpoints. 
Those fields consist only of a tower of massive vector bosons, and they do not give rise to any massless vector of scalar modes.
Also, they cannot mix with the $W$ due to the flatness of the Higgs profile, therefore their presence will not affect the previous result.
If they do couple to the Higgs, their spectrum is given by
\beq
m_n^2 = \frac{(n + 1/2 + c \alpha)^2}{L^2}\,, \quad n = 0, \pm 1, \pm 2 \dots
\eeq
where $c$ is a coefficient determined by gauge group factors.
Their contribution to $\kappagamma$ is proportional to
\beq
c\alpha \sum_{n=0}^{\infty} \frac{1}{n+1/2+c\alpha} - \frac{1}{n+1/2-c\alpha} = - \pi c\alpha \tan \pi c\alpha = - \pi^2 c^2 \alpha^2 + \mathcal{O} (\alpha^4)\,;
\eeq
therefore ($Q_X$ being the charge of the extra gauge boson)
\beq
\kappa_{\gamma \gamma } (X) \simeq - \frac{63}{16} \pi^2 Q_X^2 c^2 \frac{m_W^2}{m_{KK}^2} \sim - 0.063\, Q_X^2 c^2\cdot \left( \frac{2\, \mbox{TeV}}{m_{KK}} \right)^2\,.
\eeq
Note that it has an opposite sign compared to the $W$ tower contribution, and that it tends to be larger by a factor of $3$.

\subsubsection{Brane Higgs}

Let us first consider a bulk $SU(2)\times U(1)$ gauge symmetry, so that there is a single $W$ tower.
The spectrum is determined by the zeros of the equation (for more details, see the appendix~\ref{app:braneHiggs})
\beq \label{eq:masterWb}
f (m, \alpha) = \pi L m \tan \pi L m \; - \pi^2 \alpha^2 = 0\,,
\eeq
where $\alpha$ is again a dimensionless quantity proportional to the Higgs VEV.
The spectrum can be computed in an expansion for small $\alpha$:
\beq
m_W^2 L^2 &=& \alpha^2 \left( 1 - \frac{\pi^2}{3} \alpha^2 + \mathcal{O} (\alpha^4) \right)\,, \\
m_n^2 L^2 &=& n^2 + 2 \alpha^2 + \mathcal{O} (\alpha^4)\,. 
\eeq
In first approximation, $\alpha \sim m_W L = m_W/m_{KK}$: however higher order corrections in $\alpha$ will modify the couplings of the $W$ to the Higgs, and they must be taken into account.
Even though the spectrum cannot be calculated analytically, in the appendix we showed that
\beq
1- \frac{\alpha}{m_W} \frac{\partial m_W}{\partial \alpha} = \sum_{n=1}^\infty \frac{\alpha}{m_n} \frac{\partial m_n}{\partial \alpha} &=& 1- \frac{2 \sin (2 \pi L m_W )}{2 \pi L m_W + \sin (2 \pi L m_W)} \nonumber \\
&=& \frac{\pi^2}{3} m_W^2 L^2 + \mathcal{O} (m_W^4 L^4)\,.
\eeq
The contribution to $\kappa$ can be therefore written as:
\beq
\kappagamma (W) &=& - \frac{9}{16} \left(7 + A_W (\tau_W) \right) \left( 1-\frac{2 \sin (2 \pi L m_W )}{2 \pi L m_W + \sin (2 \pi L m_W)} \right) \nonumber\\
&\sim&  (0.003\div 0.008 )\,\cdot \left( \frac{2\, \mbox{TeV}}{m_{KK}} \right)^2
\eeq
where we have varied the $H$ mass from 115 to 150 GeV. Note that the contribution of the KK tower is the same as in the Gauge Higgs case (up to a sign): however, the total contribution is suppressed by a cancellation between the KK tower and the $W$ (it would vanish exactly in the limit of Higgs much lighter that 2 $W$ masses).

We can also consider the case of a bulk custodial symmetry, which will contain a $W_R$ gauge boson which mixes with the $W$.
In order to make it massive, one can impose Dirichlet boundary conditions on $y=0$, the opposite brane to where the Higgs is localised.
The spectrum is very similar, the only difference is to replace $L \to 2 L$ in all the equations. However, this effect is compensated by the fact that the lightest KK mode from the $W_R$ tower has mass $1/(2L)$ instead of $1/L$, therefore the result is the same as a function of the lowest KK mass.

\subsubsection{Bulk Higgs}

When the Higgs is in the bulk, it will generate a bulk mass for the gauge bosons.
The VEV therefore will shift the spectrum
\beq
m_n^2 = \frac{n^2 + \alpha^2}{L^2}\,.
\eeq
In this case, the $W$ mass ($n=0$) is proportional to the Higgs VEV ($\alpha$), so that no corrections will come from the ordinary $W$.
The contribution of the tower is proportional to the sum
\beq
\sum_{n=1}^\infty \frac{m_W^2 L^2}{n^2 + m_W^2 L^2} = \frac{\pi m_W L \coth \pi m_W L -1}{2} = \frac{\pi^2}{6} m_W^2 L^2 + \mathcal{O} (m_W^4)\,.
\eeq
The contribution to $\kappagamma$ is therefore:
\beq
\kappagamma (W) \simeq - \frac{63}{16} \frac{\pi^2}{6} m_W^2 L^2 \sim - 0.01  \left( \frac{2\, \mbox{TeV}}{m_{KK}} \right)^2\,.
\eeq
Note that the sign is different from the previous two cases, so that this contribution tends to sum up with the ordinary $W$ one.

\subsection{Gauge bosons in a warped extra dimension}

A warped extra dimension is characterised by a non-trivial metric that, in the covariant coordinates that we will be using in this paper, can be written as
\beq
d s^2 = \left( \frac{R}{z} \right)^2 (\eta_{\mu \nu} d x^\mu d x^\nu - d z^2 )\,,
\eeq
where $R$ is the curvature of the space.
Moving along the extra coordinate $z$, the unit length in 4D is rescaled, so that the natural energy scale of the model depends on the position along the warped extra coordinate.
Now, $z$ spans over an interval, but the two endpoints have a very different meaning: the brane at small $z = \epsilon$ is called the UltraViolet (UV) brane, and its typical scale represents the ultimate UV cutoff of the theory, $1/\epsilon = \Lambda$.
One can imagine that this scale is very large, say the Planck scale $M_{Pl}$.
On the other hand, at large $z = R'$ one places an InfraRed (IR) brane: its energy scale is directly related to the mass of the KK modes, so that $m_{KK} \sim 1/R'$ of order TeV.
The proportionality factor depends on the particle we are considering: for a gauge boson like the $W$, the first KK mode is at $2.4\,m_{KK}$.

The large splitting between the UV and IR scale, beyond explaining the weakness of gravitational interactions, also introduces a gap between the $W$ mass and the KK mass scale 
\beq
m_{KK} \simeq m_W \sqrt{\log \Lambda R'} \sim 6\ m_W
\eeq
for $\Lambda = M_{Pl}$ and $R' = 1$ TeV$^{-1}$.
This feature makes those models much more attractive than the flat cases, because the Higgs VEV can be closer to the IR scale.
Finally, indirect bounds will usually require $m_{KK} \geq 1$ TeV, which corresponds to a $W'$ above 2 TeV~\cite{Wcustodial,GHUwarped}, similar to the flat case.

Here we will show some features of those models, and use a numerical evaluation of the $\kappa$'s in generic models.
We focus on Gauge Higgs and IR brane Higgs models, as generic bulk Higgs models are much more complicated to deal with, both analytically and numerically~\cite{gaugephobic}.

\subsubsection{Gauge Higgs}

The spectrum of gauge bosons is determined by a complicated equation involving Bessel functions of order 1 and 0 (more details in the appendix~\ref{app:gaugehiggs}).
If we expand for large UV scale, we can get a very good approximate spectrum which depends only logarithmically on $\Lambda$.
For the $W$, assuming that it is much lighter than the KK mass, we can expand for small $m_W R' \ll 1$:
\beq
m_W^2 {R'}^2 \simeq \frac{2}{\log \Lambda R'} \sin^2 \pi \alpha \simeq  \frac{2 \pi^2}{\log \Lambda R'} \alpha^2\,,
\label{Wmass}
\eeq
where we have neglected higher order corrections in the log.
The first KK mode will be given by the zeros of Bessel functions and one finds $m_{W'} \sim 2.4/R'$.
Note also that this expression fixes $\alpha$ as a function of the KK scale $R'$.
Contrary to the flat case, the $W$ mass is not linear in the Higgs VEV, so that there will be corrections coming from the deviations from the SM coupling to the Higgs.
Those corrections are proportional to
\beq
\frac{\alpha}{m_W} \frac{\partial m_W}{\partial \alpha} -1 \simeq \pi \alpha \cot \pi \alpha\, -1 \simeq - \frac{m_W^2 {R'}^2}{6} \log \Lambda R' \sim -0.037 \left( \frac{1/R'}{1\mbox{TeV}} \right)^2\,. 
\eeq
Note that one can use the same trick that we used in the flat brane Higgs (see appendix ~\ref{app:braneHiggs}) to calculate this quantity exactly: however, for our purpose, this approximate result is more than sufficient, and we can check that sub-leading terms will be suppressed by a log compared to this result:
\beq
\frac{\alpha}{m_W} \frac{\partial m_W}{\partial \alpha} -1 = - \frac{m_W^2 {R'}^2}{6} \log \Lambda R' \left( 1 - \frac{9}{4} \frac{1}{\log \Lambda R'} + \frac{3}{2} \frac{1}{\log^2 \Lambda R'} \right) + \mathcal{O} (m_W^4 {R'}^4)\,.
\eeq
The contribution to $\kappagamma$ is
\beq
\kappagamma (W) \sim \frac{9}{16} (-0.037) A_W (\tau_W) \sim (0.17 \div 0.21 )\cdot \left( \frac{1/R'}{1\mbox{TeV}} \right)^2 
\eeq
where $m_H = 115 \div 150$ GeV.

One can also numerically compute the KK tower contribution to the $\kappa$'s and find (for a $W$ tower):
\beq
\kappagamma (W_{KK}) \sim 0.009\, \cdot\, \left( \frac{1\, \mbox{TeV}}{1/R'} \right)^2
\eeq
for $\Lambda = M_{Pl}$.
This contribution is much smaller that the contribution of the $W$.

\subsubsection{Brane Higgs}

Expanding for large $\Lambda$ and small $m_W R' \ll 1$, the mass of the $W$ is:
\beq
m_W^2 {R'}^2 = \frac{\alpha^2}{(1+\alpha^2/2) \log \Lambda R'} + \dots 
\eeq
Similarly to the Gauge Higgs case, the coupling of the $W$ to the Higgs will receive corrections, and the contribution to the $\kappa$ will be proportional to
\beq
\frac{\alpha}{m_W} \frac{\partial m_W}{\partial \alpha} -1 = - \frac{m_W^2 {R'}^2}{2} \log \Lambda R' \left( 1 - 2 \frac{1}{\log \Lambda R'} +  \frac{1}{\log^2 \Lambda R'} \right) + \mathcal{O} (m_W^4 {R'}^4)\,.
\eeq
We also numerically computed the contribution of the KK tower, and, as in the flat case, the following relation holds:
\beq
\sum_{n=1}^\infty \frac{\alpha}{m_n} \frac{\partial m_n}{\partial \alpha} = 1- \frac{\alpha}{m_W} \frac{\partial m_W}{\partial \alpha}\,. 
\eeq
Numerically:
\beq
\kappagamma (W) \sim - \frac{9}{16} (0.11) \left(7+ A_W (\tau_W) \right) \sim (0.07 \div 0.18 )\cdot \left( \frac{1/R'}{1\mbox{TeV}} \right)^2 
\eeq
where $m_H = 115 \div 150$ GeV.
As in the flat case, the two contributions tend to cancel each other in the light Higgs limit.

\subsection{Bulk fermions in a flat extra dimension}

Bulk fermions are easier to analyse because the basic structure is common to all kind of Higgs models: we always need two bulk fermions, a doublet and a singlet of SU(2), that couple via the Higgs (either in the bulk or on the brane).
In Gauge Higgs, those fields are in the same representation of the extended bulk gauge symmetry, while in other models they can be independent fields.
Both in Gauge Higgs and in the brane Higgs case, the Higgs appears in the boundary conditions, which have the same form in the two cases: the reason is that one can use a gauge transformation to remove the Gauge Higgs VEV from the bulk equations of motion, as explained in more detail in the appendix~\ref{app:fermions}.
The only difference is that the boundary conditions depend differently on the Higgs VEV.
In the following we will use the notation of the Gauge Higgs models, where the Higgs VEV enters via trigonometric functions of a dimensionless parameter $\beta$.
Note that the $\beta$ parameter is different in general from the one for the gauge bosons $\alpha$, due to either gauge group factors or Yukawa couplings.
In Gauge Higgs, both $\beta$ and $\alpha$ are proportional to the Higgs VEV.
In the brane Higgs case, we can also define a fictitious $\beta$ parameter that is related to the actual brane Higgs VEV $V$ (see the appendix \ref{app:fermions} for more details) as
\beq
\tan \pi \beta = y V\,,
\eeq
where $y$ is an effective Yukawa coupling.
The spectrum will be the same in the two cases, as a function of $\beta$, however the couplings to the Higgs are different.
The results in the brane Higgs case are equal to the Gauge Higgs case, up to a correction factor
\beq
\frac{V}{\beta} \frac{\partial \beta}{\partial V} = \frac{\sin \pi \beta\; \cos \pi \beta}{\pi \beta}\,.
\eeq
This factor takes into account the non linear relation between $\beta$ and the brane Higgs VEV $V$.

In the Bulk Higgs case, the spectrum is different: the calculation is more complicated in the case of a generic Higgs profile, and we will only study the case of a constant Higgs VEV in a flat extra dimension, which is relevant for the UED model.

\subsubsection{Bulk fermions}

The first spectrum we will consider is the following:
\beq
m_n^2 = M^2 + \frac{(n + \beta)^2}{L^2}\,, \quad n = 0, \pm 1, \pm 2 \dots
\eeq
This spectrum can arise in many scenarios: in Gauge Higgs models, one can generate the masses of light fermions by using two copies of bulk fermions with opposite boundary conditions, and connected by a bulk mass term $M$~\cite{serone}.
The light fermions are localised degrees of freedom that can mix with the massive bulk fields via localised mass terms: the electroweak symmetry breaking is mediated to the localised fields by the massive ones, like in the Froggatt-Nielsen model.
In this case, the smallness of the light mass can be achieved either by a small mixing, or by a large bulk mass $M$.

For $M=0$
\beq
m_n^2 =  \frac{(n + \beta)^2}{L^2}\,, \quad n = 0, \pm 1, \pm 2 \dots
\eeq
and we have a light mode with mass
\beq
m_f = \frac{\beta}{L}\,.
\eeq
In Gauge Higgs models, this can be identified with the top, whose mass is of order the electroweak scale: a $m_{top} \neq m_W$ can be obtain by using a large representation for the field containing the top~\cite{GHUflat1}, or by an explicit breaking of Lorentz invariance~\cite{GHUflat2}. 
As we will shortly see, warped geometry automatically solves this problem.
In brane Higgs models, all fermion masses can be generated in this way, because the relation between $\beta$ and the Higgs VEV depends on Yukawa couplings and each field will feel a different effective $\beta$ parameter.
Finally, this spectrum will be generated also by a Bulk Higgs with a flat profile, where $\beta$ is proportional to the Higgs VEV.
For more detail about the spectra, see the appendix~\ref{app:fermions}.

The $\kappa$'s will be proportional to the sum
\beq
\sum_n \frac{\beta}{m_n} \frac{\partial m_n}{\partial \beta} &=& \frac{\beta^2}{(M L)^2 + \beta^2} + \beta \sum_{n=1}^{\infty} \frac{n+\beta}{(M L)^2 + (n + \beta)^2} -  \frac{n-\beta}{(M L)^2 + (n - \beta)^2} = \nonumber \\
& = & \frac{\pi \beta \sin (2 \pi \beta)}{\cosh (2 \pi M L) - \cos (2 \pi \beta)}\,, \label{eq:flatpiupiu}
\eeq
with the proportionality coefficient determined by the quantum numbers of the 5D field (charge and colour), and a correction factor in the brane Higgs case.
For large $M L$, this contribution is exponentially suppressed: the mass of the localised fermions is also suppressed by $\exp (- \pi M L)$, therefore we find
\beq
 \kappa \sim \frac{m_f^2}{m_{KK}^2}\,.
\eeq
In this class of models, one can safely neglect the contribution of the light fermion towers.

For the top in Gauge Higgs, or in the case of brane Higgs and Bulk Higgs models ($M=0$), $\beta = m_f L$ and
\beq
\sum_n \frac{\beta}{m_n} \frac{\partial m_n}{\partial \beta}  = \pi \beta \cot \pi \beta\,. \label{eq:flatpiupiu0}
\eeq
The contribution of the top tower is:
\beq
\kappa_{\gamma \gamma} = \kappa_{gg} = \left( \pi \frac{m_t}{m_{KK}} \cot \left( \pi \frac{m_t}{m_{KK}}\right) -1 \right) \simeq - \frac{\pi^2}{3} \frac{m_t^2}{m_{KK}^2} \sim - 0.025  \left( \frac{2\, \mbox{TeV}}{m_{KK}} \right)^2\ \,,
\eeq
where we have subtracted the top contribution.
In the brane Higgs case:
\beq
\kappa_{\gamma \gamma} = \kappa_{gg} = \left( \cos^2 \left( \pi \frac{m_t}{m_{KK}}\right) -1 \right) \simeq - \pi^2 \frac{m_t^2}{m_{KK}^2} \sim - 0.075  \left( \frac{2\, \mbox{TeV}}{m_{KK}} \right)^2\ \,.
\eeq
The contribution from other light fermion towers are negligible, as they will be proportional to the light fermion mass squared.

For completeness, let us also report the contribution from a tower of states with twisted boundary conditions, which may be present in models with large bulk representation and have a spectrum
\beq
m_n^2 = M^2 + \frac{(n +1/2 + \beta)^2}{L^2}\,, \quad n = 0, \pm 1, \pm 2 \dots
\eeq
In this case:
\beq
\sum_n \frac{\beta}{m_n} \frac{\partial m_n}{\partial \beta} &=& \beta \sum_{n=0}^{\infty} \frac{n+1/2+\beta}{(M L)^2 + (n +1/2 + \beta)^2} -  \frac{n+1/2-\beta}{(M L)^2 + (n +1/2- \beta)^2} = \nonumber\\
& = & - \frac{\pi \beta \sin (2 \pi \beta)}{\cosh (2 \pi M L) + \cos (2 \pi \beta)}\,. \label{eq:flatpiumeno}
\eeq
This contribution is also suppressed for large bulk masses.
In the $M \to 0$ limit we get:
\beq
\sum_n \frac{\beta}{m_n} \frac{\partial m_n}{\partial \beta} = - \pi \beta \tan \pi \beta  \simeq - \pi^2 \beta^2 \label{eq:flatpiumeno0}\,.
\eeq
The contribution to the $\kappa$'s is
\beq
\kappaglu (\mbox{twisted}) &=& - \left\{ \begin{array}{cc}
 \pi \beta \tan \pi \beta \\
\sin^2 \pi \beta
\end{array} \right\}
\sim - 0.075\; \frac{m_f^2}{m_t^2} \; \left( \frac{2\, \mbox{TeV}}{m_{KK}} \right)^2\,, \\
\kappagamma (\mbox{twisted}) &=& \frac{9}{4} Q_f^2\; \kappaglu\,,
\eeq
for a colour-triplet with $m_f L = \beta$.
The two results in the brackets correspond to GH/BH (up) and bH (down), and they give the same contribution in the small $\beta$ limit.

\subsubsection{Bulk fermion in UED models}

In this case the spectrum is (similarly to the gauge case)
\beq
m_n^2 = \frac{n^2 + \beta^2}{L^2}\,, \quad n = 0, 1, 2 \dots 
\eeq
with $\beta$ proportional to the Higgs VEV via the bulk Yukawa coupling: for any SM fermion, $\beta = m_f L$, therefore only the top quark is relevant.
The contribution to the amplitude is the same as for the gauge bosons, so that the top KK tower gives
\beq
\kappagamma (top) = \kappaglu (top) \simeq \frac{\pi^2}{6} m_t^2 L^2 \sim 0.01 \left(\frac{2 \mbox{TeV}}{m_{KK}} \right)^2\,.
\eeq
Note that this contribution has an opposite sign compared to the GH/bH cases, and, in $\kappagamma$, it tends to cancel the contribution of the $W$ tower.

\subsubsection{Odd bulk masses: fermions in models of flavour}

Lorentz invariance in 5D allows to write down a mass term for a single 5D fermion $\tilde M$: this mass term is however forbidden in orbifold models, because the two components of the 5D fermions have opposite parities (unless the mass has an odd profile).
A model defined on an interval is less constrained as it allows for the presence of such masses.
Those odd masses have a very important phenomenological feature~\cite{localization}: the zero mode of the 5D fermion is chiral, therefore $\tilde M$ cannot give it a mass!
Its effect is to exponentially localise the zero mode, and therefore modify the overlap with other fields, in particular with the Higgs (either bulk or localised).
This feature can be used in a variety of models to generate the hierarchies in the fermion masses using order 1 Yukawas and bulk masses for all fermion fields!
This is an alternative mechanism to generate light fermions in GH models, where the Yukawa couplings are equal due to gauge invariance, but it can also be used in bH and BH models.
There is however a crucial difference between the two: in GH models the bulk masses are the same for the two SM fields that couple to the Higgs, because they come from the same bulk multiplet, while in bH/BH models they can be different.
As we will see, this has dramatic consequences for the Higgs phenomenology.

Here we will focus on the GH and bH cases: the Higgs VEV enters via a dimensionless parameter $\beta$, and we will define
\beq
m_f = \beta/L\,.
\eeq
As a reference, we will assume $m_f = m_{top}$, but this may not be the case in all models.
The equation determining the spectrum is more complicated than in the previous case, therefore we will limit ourselves to an expansion for small $\beta$.
In the GH case, the odd masses are the same for the two bulk fields, $\tilde M$.
Expanding for $m L \ll \tilde  M L$, we can calculate the mass of the light mode, which would be identified with the SM fermion:
\beq
m_l^2 = \frac{2 \tilde M^2 \sin^2 \pi \beta}{\cosh (2 \pi \tilde M L) - \cos (2 \pi \beta)} \simeq \frac{2\tilde M^2 L^2 \pi^2}{\sinh^2 \pi \tilde M L} m_f^2\,.
\eeq
It is clear from this formula that the light mode mass is suppressed by $\exp (- \pi \tilde M L)$ compared to the Higgs VEV $\beta$.
Therefore, a $\tilde M L \sim \mathcal{O} (1)$ can explain the lightness of the fermions in the SM.
The spectrum of the heavy modes, $m_n > \tilde M$, is more complicated:
\beq
m_n^2 L^2= \tilde M^2 L^2 + n^2 \pm \frac{2 n^2}{\sqrt{n^2 + \tilde M^2 L^2}} \beta + \frac{n^4 + 3 \tilde M^2 L^2 n^2}{(n^2 + \tilde M^2 L ^2)^2}\beta^2 + \mathcal{O} (\beta^3) \,.
\eeq 
The couplings of each mode to the Higgs are large, however like in the previous case the modes have a different sign in the coupling.
The sum, therefore, gives at leading order in $\beta$:
\begin{multline}
\sum_n \frac{\beta}{m_n} \frac{\partial m_n}{\partial \beta} = - 2 \beta^2 \sum_{n=1}^{\infty} \frac{n^4 - 3 \tilde M^2 L^2 n^2}{(n^2 +\tilde M^2 L ^2)^2} = \\
 - \frac{\pi^2 \beta^2}{\sinh^2 \pi \tilde M L} \left( \frac{\pi \tilde M L}{\tanh \pi \tilde M L} - 1 \right) \simeq - \frac{m_l^2}{2 \tilde M^2}  \left( \frac{\pi \tilde M L}{\tanh \pi \tilde M L} - 1 \right)\,.
\end{multline}
We find again that the tower of light modes does not contribute significantly to the widths: the only contribution will come from the top tower ($\tilde M \to 0$) which is approximate by the result in the previous subsection.
Note again that this exponential suppression comes from a non trivial cancellation between modes.

In models with brane Yukawa couplings, the fields containing the SU(2) doublet and singlet are not necessarily the same, so they can have different bulk masses, $\tilde M_L$ and $\tilde M_R$.
In this case, the spectra of the two bulk fermions are different, the two KK towers are not degenerate in the $\beta \to 0$ limit and there are no cancellations between modes.
As we will see, the spectra are degenerate if $\tilde M_R = - \tilde M_L$, however, even in this case, the cancellation between modes that we observe in the GH case does not occur.
In conclusion, in models of this kind, the contribution of the KK tower of light modes can be large, as it is proportional to the 5D Yukawa coupling and not to the effective light-mode Yukawa (light fermion mass).
As an example, we can study the latter case $\tilde M_L = - \tilde M_R =\tilde M$, where some simple analytical results can be obtained.
The zero mode mass is, at leading order,
\beq
m_l \simeq 4 \tilde M e^{- 2 \pi \tilde M L} \sin \pi \beta\,,
\eeq
suppressed by $\exp (- 2 \pi \tilde M L)$.
As before, we can compute the approximate KK spectrum for small $\beta$
\begin{multline}
m_n^2 L^2= \tilde M^2 L^2 + n^2 \pm \frac{2 n^2}{\sqrt{n^2 + \tilde M^2 L^2}} \beta + \\
 \frac{(1-2 \pi\tilde M L) n^4  + (3 - 2 \pi \tilde M L) \tilde M^2 L^2 n^2}{(n^2 + \tilde M^2 L ^2)^2} \beta^2 + \mathcal{O} (\beta^3) \,,
\end{multline} 
which differs to the $\tilde M_L = \tilde M_R$ case only at order $\beta^2$, and the sum over the massive modes (the SM fermion is not included)
\begin{multline}
\sum_n \frac{\beta}{m_n} \frac{\partial m_n}{\partial \beta}  = - 2 \beta^2 \sum_{n=1}^{\infty} \frac{(1+2 \pi \tilde M L) n^4 - (3 - 2 \pi \tilde M L)  \tilde M^2 L^2 n^2}{(n^2 + \tilde M^2 L ^2)^2} = \\
 - \frac{\pi^2 \beta^2}{4 \sinh^3 \pi \tilde M L} \left( \cosh (3 \pi \tilde M L) + (4 \pi \tilde M L-1 ) \cosh (\pi \tilde M L) - 4 (\pi \tilde M L +1) \sinh (\pi \tilde M L) \right) \\
\simeq - \pi^2 \beta^2 \sim - 0.075 \left( \frac{2 \mbox{TeV}}{m_{KK}} \right)^2 \left( \frac{m_f}{m_{top}} \right)^2\,. \phantom{pippopippopippo} 
\end{multline}
The result is not very sensitive to the precise value of the bulk masses, even in the case of $\tilde M_L \neq \tilde M_R$ (we checked this numerically).
Moreover, corrections from the non-linear relation between $\beta$ and the Higgs VEV (the same multiplicative factor as in the previous section) will only affect this result at higher orders in $\beta$, while the light mode is negligible. 
The contribution of the top tower will be the same as in the massless case and, at leading order, it also gives $- \pi^2 \beta^2$.
For a model with this flavour structure, contributions of the light fermion and top towers are:
\beq
\kappagamma = \kappaglu \simeq 6 ( - \pi^2 \beta^2 ) \sim - 0.45  \left( \frac{2 \mbox{TeV}}{m_{KK}} \right)^2\,,
\eeq
where the factor of 6 takes into account 3 complete SM generations, and we assumed that all the Yukawa couplings are of the same order (as the top one).

\subsection{Fermions in a warped extra dimension}

The localisation mechanism in warped extra dimension is much more effective than in the flat case: the reason is that the localisation is exponential with the large number $\Lambda R'$.
The geometry itself generates two hierarchical mass scales: the UV cutoff $\Lambda$ on the UV brane and the KK scale $1/R'$ on the IR brane.
Here we will use the usual notation to call $c$ the odd bulk mass in units of the curvature, $c = \tilde M R$.
A left-handed (right-handed) zero mode is localised on the  UV brane for $c>1/2$ ($c<-1/2$) and IR brane for $c<1/2$ ($c>-1/2$)~\cite{bulkmass}.

GH models are characterised by the same odd mass $c$ for the two fields that couple to the Higgs, because they are part of the same bulk multiplet.
Like in the gauge boson case, we can expand for large UV cutoff, however in the fermionic case the expansion is more complicate and depends on the value of the bulk mass $c$.
For $-1/2 < c < 1/2$ (when both zero modes are localised on the IR brane), the mass of the light mode is
\beq
m_f R' \simeq \sqrt{1-4 c^2}\; \pi \beta \left( 1 - \frac{3+4 c^2}{9-4 c^2} \pi^2 \beta^2 + \mathcal{O} (\beta^4) \right)\,.
\eeq
The mass is not suppressed compared to the Higgs VEV $\beta$; notice also that the log suppression between the $W$ mass and $\beta$ is not present here, therefore one can fit the top mass without using a large representation (therefore, $\beta = \alpha$ is acceptable)!
The mass does not depend linearly on $\beta$, thus the coupling with the Higgs receives corrections compared to the SM value, that will contribute to the $\kappa$'s.
For a fermion with the same quantum numbers of the top (and in the light-Higgs approximation):
\beq
\kappagamma (t)= \kappaglu (t) = \frac{\beta}{m_f} \frac{\partial m_f}{\partial \beta} -1 \simeq - \frac{32}{3} \frac{c^2}{(9 - 4 c^2)(1 - 4 c^2)} m_f^2 {R'}\,.
\eeq
We also calculated this contribution exactly, and verified that this approximation is good for $|c| < 0.4$ at a few percent level.
The $\kappa$'s vanish for $c\to 0$: in fact, in this limit the Bessel functions reduce to sines and cosines and we recover the flat case result where the light fermion mass is linear in the Higgs VEV.
The coupling of the Higgs to the KK modes is also large.
In summary, for $-1/2 < c < 1/2$, the light mass is un-suppressed compared to the Higgs VEV and the contribution of the tower to the $\kappa$'s is sizable.
Numerically we found
\beq \label{eq:kappacmhalf}
\kappagamma (t_{KK}) = \kappaglu (t_{KK}) \sim - \frac{1}{3} m_f^2 {R'}^2
\eeq
for a fermion tower with the same quantum numbers of the top.
This contribution will sum with the one coming from the top; notice that it is very similar to the flat case result (factor of $1/3$).
For the top quark in GH (with $1/R = 1$ TeV, and $\beta = \alpha$ fixed by the $W$ mass), we need $c\sim 0.43$: numerically
\beq
\kappagamma (top) = \kappaglu (top) \sim (-0.029) + (-0.011) = -0.04 \cdot \left( \frac{1\mbox{TeV}}{1/R} \right)^2 \,.
\eeq
Note finally that this result can also be generalised to the brane-Higgs case with equal masses, taking into account the correction mentioned in the previous section, which is the same independently on the geometry.

\begin{figure}[t]
\centering
\includegraphics[width=0.5\textwidth]{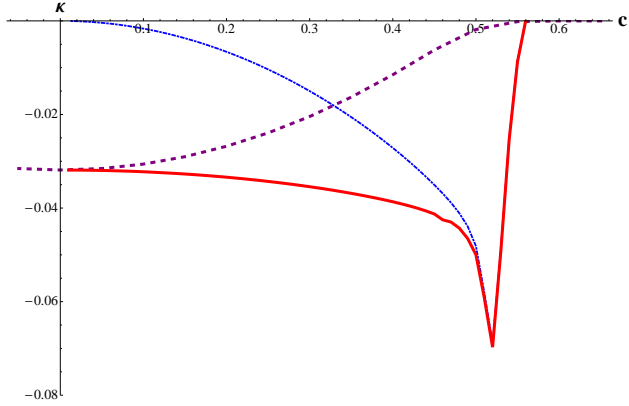}
\caption{\footnotesize Contribution to $\kappa$ from a bulk fermion in GHU as a function of the bulk mass $c$, $1/R = 1$ TeV, and $m_H = 130$ GeV. \textcolor{blue}{The dotted line} represents the contribution of the light mode (deviation from the SM one), \textcolor{purple}{the dashed line} is for the KK tower and \textcolor{red}{the thick one} corresponds to the sum of both. }
\label{fig:fermionsGHU}
\end{figure}

For $c>1/2$ and $c<-1/2$ the two zero modes are localised on different endpoints, and the light mass is suppressed:
\beq
m_f R'  \simeq \left( \frac{1}{\Lambda R'} \right)^{|c|-\frac{1}{2}}\;\sqrt{4 c^2 -1}\, \sin \pi \beta\,.
\eeq
In this case the coupling of the KK modes to the Higgs is also suppressed by:
\beq
\left( \frac{1}{2 \Lambda R'} \right)^{2 |c| -1} \sim \frac{m_f^2}{m_{KK}^2}\,.
\eeq
The contribution of a light fermion KK tower is negligible~\footnote{This result agrees with Ref.~\cite{falkowski}.}, however, contrary to the flat case, there is no cancellation involved and each KK mode coupling is suppressed by the light fermion mass.
In figure~\ref{fig:fermionsGHU} we computed numerically the contribution of a bulk fermion as a function of the bulk mass $c$ ($\beta = \alpha$ fits the $W$ mass and $1/R' = 1$ TeV).
The contribution of the light fermion is the deviation from a SM fermion of the same mass: it vanishes for $c=0$, grows towards $c=1/2$ reaching the value calculated for the $W$, and then goes down due to the decrease in the SM amplitude for a light fermion.
The contribution of the KK modes, on the other hand, decreases for large $c$.
The total contribution is almost constant for $c<1/2$, then reaches a peak when the light fermion is at the Higgs decay threshold (in the plot, $m_H = 130$ GeV), and then goes rapidly to zero.

It is straightforward to understand the suppression if we analyse in detail the structure of the wave functions: let us consider first the doublet, which contains a left-handed zero mode.
The wave functions, after applying the UV boundary conditions, are
\beq
\chi_L &=& A_L z^{5/2} \left[ J_{-c+\frac{1}{2}} (p/\Lambda)  J_{c+\frac{1}{2}} (p z) +  J_{c-\frac{1}{2}} (p/\Lambda)  J_{-c-\frac{1}{2}} (p z) \right] \\
\psi_L &=& A_L z^{5/2} \left[ J_{-c+\frac{1}{2}} (p/\Lambda)  J_{c-\frac{1}{2}} (p z) -  J_{c-\frac{1}{2}} (p/\Lambda)  J_{-c+\frac{1}{2}} (p z) \right]
\eeq
where $\chi$ ($\psi$) is the left-handed (right-handed) wave function component.
The expansion for small $p/\Lambda$ depends on the value of c ($J_\nu (x) \sim x^\nu$ for small $x$):
\beq
\chi_L \simeq \left\{ \begin{array}{lc}
J_{c+\frac{1}{2}} (p z) & \mbox{for}\, c>1/2 \\
J_{-c-\frac{1}{2}} (p z) & \mbox{for}\, c<1/2
\end{array} \right. \qquad
\psi_L \simeq \left\{ \begin{array}{lc}
J_{c-\frac{1}{2}} (p z) & \mbox{for}\, c>1/2 \\
J_{-c+\frac{1}{2}} (p z) & \mbox{for}\, c<1/2
\end{array} \right.
\eeq
plus corrections suppressed by $\left( \frac{p}{\Lambda} \right)^{|2c-1|}$.
For the singlet field, that contains the right-handed zero mode
\beq
\chi_R \simeq \left\{ \begin{array}{lc}
J_{c+\frac{1}{2}} (p z) & \mbox{for}\, c>-1/2 \\
J_{-c-\frac{1}{2}} (p z) & \mbox{for}\, c<-1/2
\end{array} \right. \qquad
\psi_R \simeq \left\{ \begin{array}{lc}
J_{c-\frac{1}{2}} (p z) & \mbox{for}\, c>-1/2 \\
J_{-c+\frac{1}{2}} (p z) & \mbox{for}\, c<-1/2
\end{array} \right.
\eeq
plus corrections suppressed by $\left( \frac{p}{\Lambda} \right)^{|2c+1|}$.
The IR boundary conditions are:
\beq
\psi_L \cos \pi \beta + i \psi_R \sin \pi \beta &=& 0\,,\\
\chi_R \cos \pi \beta + i \chi_L \sin \pi \beta &=& 0\,.
\eeq
It is clear that if the wave functions are proportional to each other, $\psi_L \propto \psi_R$ and $\chi_L \propto \chi_R$, the $\beta$ dependence drops out from the equations: this is indeed the case at leading order for $c>1/2$ and $c<-1/2$.
In this case the Higgs VEV will affect the spectrum only via a suppressed contribution.

We can apply the same discussion to the generic brane localised Higgs: in this case, there are two different bulk masses $c_L$ and $c_R$.
Unless $c_L = c_R$, the wave functions are different and cannot be proportional to each other, therefore the coupling of the Higgs will be sizable even though the zero mode mass is suppressed due to the localisation of its wave functions.
As in the flat case, the towers of light modes will give a large contribution to the $\kappa$'s.
It can be calculated numerically and we found that for $c_L > 1/2$ and $c_R <-1/2$ it can be approximated by
\beq
\kappagamma = \kappaglu \simeq - \pi^2 \beta^2 \sim - 0.12\; \cdot\; \left( \frac{1\, \mbox{TeV}}{1/R'} \right)^2
\eeq
for a fermion tower with the same quantum numbers of the top, $\Lambda = M_{Pl}$ and using $\beta = \alpha$ that fits the $W$ mass.
Like in the flat case, the contribution of the KK towers of the light fermions is very large.
To conclude, we can quote the number for a realistic quark and lepton spectrum. We use $c_L^{top} = 0.37$, $c_R^{top}=0$, $c_R^{bot} = -0.55$:
\beq
\kappagamma ({\rm fermions}) \simeq \kappaglu ({\rm fermions}) \sim -0.65\, \cdot \, \left( \frac{1\,\mbox{TeV}}{1/R'} \right)^2\,.
\eeq

\section{Numerical results}
\label{sec:numres}
\setcounter{equation}{0}

\begin{figure}[!t]
\begin{center}
\includegraphics[width=1\textwidth]{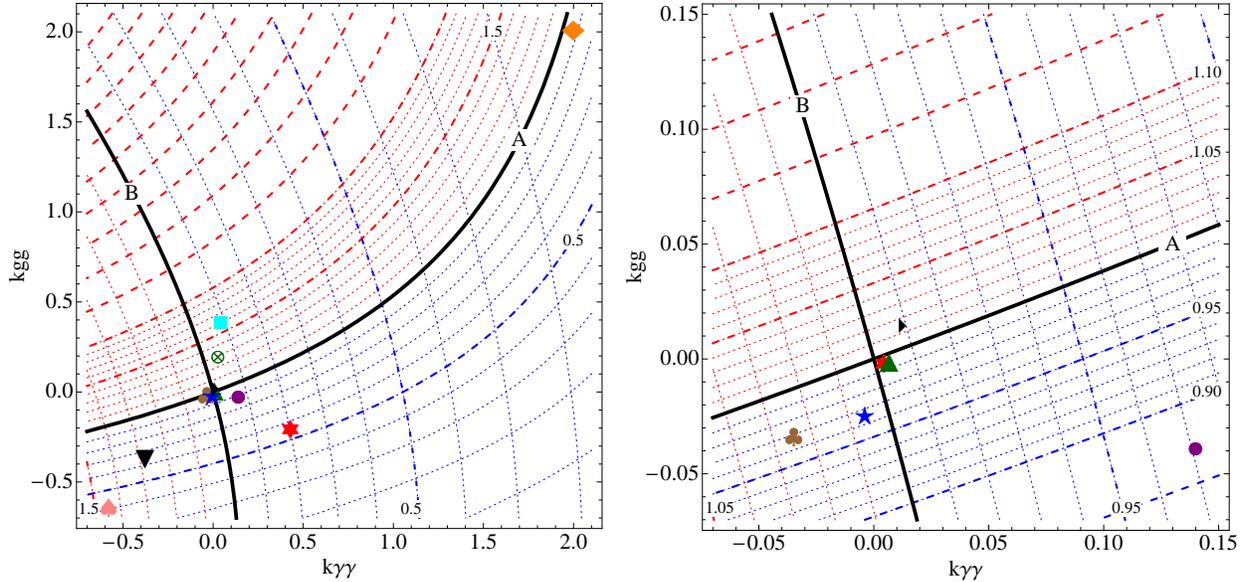}
\end{center}
\caption{\footnotesize  $\kappagamma$ and $\kappaglu$ at the LHC for a light Higgs ($m_H = 120$ GeV).
The two solid lines correspond to the SM values of the inclusive $\gamma \gamma$ channel ({\bf A}), and the vector boson fusion production channel ({\bf B}).
On the left panel, the dashed lines are spaced by $0.5$, while the dotted ones by $0.1$. On the right, we zoomed near the SM point.}
\label{fig:light}
\end{figure}

\begin{figure}[!t]
\begin{center}
\includegraphics[width=1\textwidth]{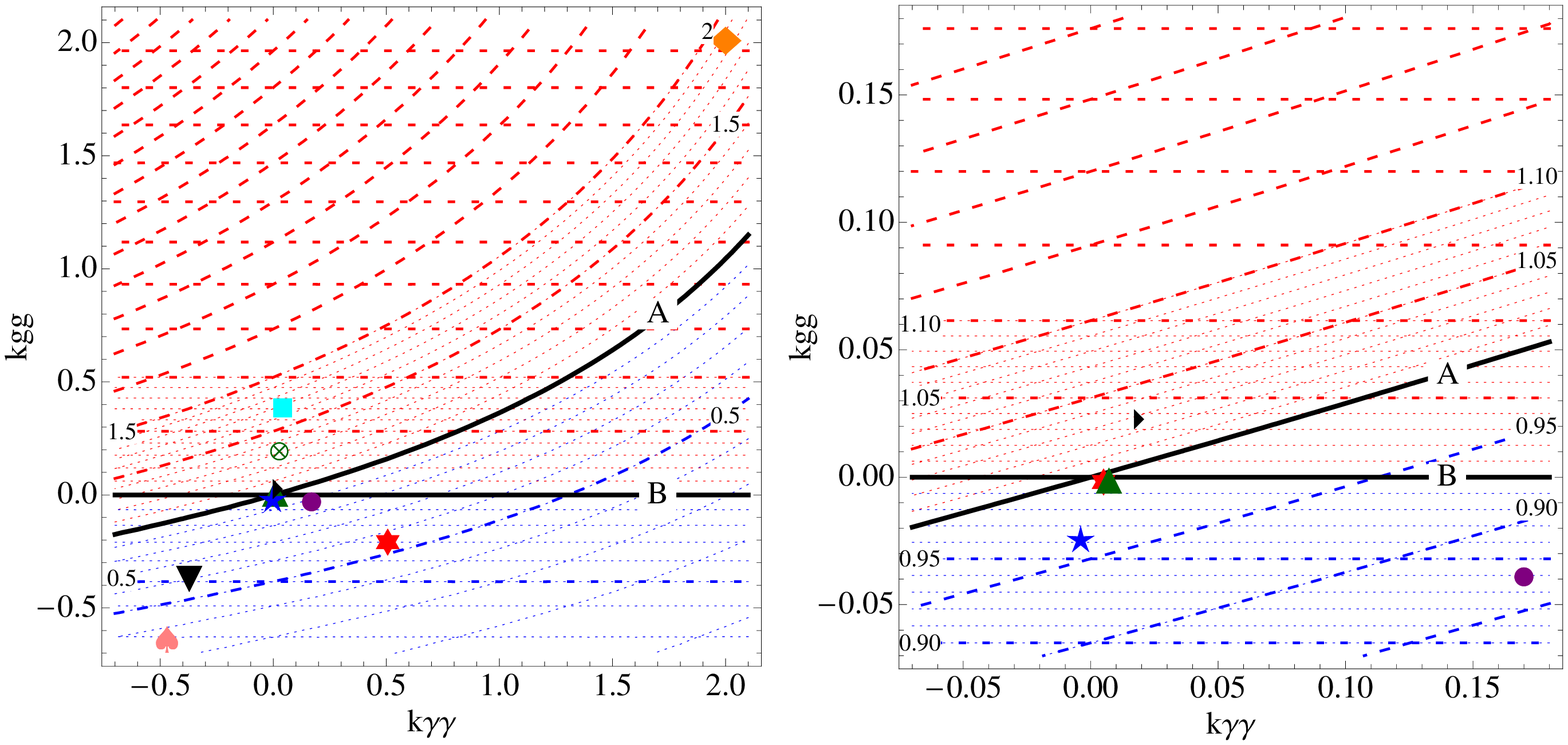}
\end{center}
\caption{\footnotesize    $\kappagamma$ and $\kappaglu$ at the LHC for a Higgs near the $WW$ threshold ($m_H = 150$ GeV).
The two solid lines correspond to the SM values of the inclusive $\gamma \gamma$ channel ({\bf A}), and the inclusive $V^* V$ channel ($V=W,Z$) ({\bf B}).
On the left panel, the dashed lines are spaced by $0.5$, while the dotted ones by $0.1$. On the right, we zoomed near the SM point.
}
\label{fig:heavy}
\end{figure}

In this section we present exact numerical results for the models we considered in the previous two sections: in all cases, the analytic formulae are a very good approximation.
We considered the following models:
\begin{itemize}
\item[-] [\textcolor{orange}{$\blacklozenge$}] a fourth generation (the result is independent on the masses and Yukawa couplings);
\item[-] [\textcolor{marron}{ $\clubsuit$}] supersymmetry in the MSSM golden region: we only included the contribution of the stops with the spectrum given by the benchmark point in~\cite{Perelstein:2007nx}. In this case the result is very sensitive to the parameters in the superpotential and in the susy breaking terms, therefore the general MSSM will cover a region of the parameter space;
\item[-] [\textcolor{vertfonce}{$\blacktriangle$}] Simplest Little Higgs, the result scales with the $W'$ mass (in the plots, $m_{W'}=2$ TeV); 
\item[-] [\textcolor{red}{$\ast$}] Littlest Higgs, the result scales with the symmetry breaking scale $f$ and has a mild dependence on the triplet VEV $x$ (we set $x=0$): for a model with T-parity we use $f=500$ GeV, without T parity $f=5$ TeV;
\item[-] [\textcolor{cyan}{$\blacksquare$}] colour octet model, the result depends on 2 free parameters: for illustration we use in the plots $X_1 = 1/9$ and $X_2 = 1/36$ (see Section~\ref{sec:coloctet});
\item[-] [$\blacktriangleright$] Lee-Wick Standard Model, the result scales with the LW Higgs mass: in the plots we set it to $1$ TeV for illustration;
\item[-] [\textcolor{vertfonce}{$\otimes$}] Universal Extra Dimension model~\cite{UED}, where only the top and $W$ resonances contribute and the result scales with the size of the extra dimension: here we set $m_{KK} = 500$ GeV close to the experimental bound;
\item[-] [\textcolor{blue}{$\bigstar$}] the model of Gauge Higgs unification in flat space in Ref.~\cite{GHUflat2}, where only the $W$ and top towers contribute ($\beta = m_t L$), with the first $W$ resonance at 2 TeV;
\item[-] [\textcolor{purple}{$\bullet$}] the Minimal Composite Higgs~\cite{GHUwarped} (Gauge Higgs unification in warped space) with the IR brane at $1/R' = 1$ TeV: only $W$ and top towers contribute significantly. The point only depends on the overall scale of the KK masses, as the other parameters are fixed by the $W$ and top masses;
\item[-] [\textcolor{gray}{$\blacktriangledown$}] a flat ($W'$ at 2 TeV) and [\textcolor{pink}{$\spadesuit$}] warped ($1/R'$ at 1 TeV) version of brane Higgs models, in both cases the hierarchy in the fermionic spectrum is explained by the localisation, and all light fermion towers contribute. Notwithstanding the many parameters in the fermion sector, the result only depends on the overall scale of the KK masses.
\end{itemize}
In the numerical results, the value of the mass of the new particles is at or around the lower bound given by precision electroweak tests; for larger masses, the contribution scales like the inverse squared mass (with the exception of the fourth generation).
Note that in many cases, the result only depends on one mass scale, and is insensitive to other free parameters present in the model: for example, in extra dimensional models with flavour, the final result does not depend on the precise localisation pattern of the bulk fields.
Therefore, changing the parameters of the model can only move the point towards the origin by increasing such mass scale (except for supersymmetry and the colour octet model, where a wide region of the parameter space may be covered).
The models are displayed in Figures~\ref{fig:light} and~\ref{fig:heavy}: different classes of models point in different quadrants of the parameter space.
Therefore, if we could measure experimentally the two parameters, depending on the accuracy of the measurements, we may be able to distinguish between models and have an hint of what kind of mechanism lies behind the breaking of the electroweak symmetry.
The direct discovery of the new particles would then be a confirmation of the model.
The complementarity between the two measurements is crucial, because this indirect probe is sensitive to the quantum numbers and couplings to the Higgs of the new particles.
This information is hardly accessible at the LHC, except in some special cases.
It is crucial to understand the reach and discrimination power of the LHC in this parameter space.

The LHC will surely be able to measure the inclusive cross section $\sigma ( p p \to H \to \gamma \gamma)$, as this is one of the golden channels for the discovery of a light Higgs.
For an integrated luminosity of 10 fb$^{-1}$ we can expect a 10\% accuracy with respect to the Standard Model one~\cite{LHCreach}.
We plotted the inclusive cross section normalised by the SM value in the $\kappagamma$--$\kappaglu$ parameter space for a light Higgs ($m_H = 120$ GeV) in Figure~\ref{fig:light} and for a Higgs near the $VV$-threshold ($m_H = 150$ GeV) in Figure~\ref{fig:heavy}: many models lie very far from such line, and a 10\% measurement would allow to probe new physics masses up to few TeV in some cases.
Note that many of the models we studied predict a reduction of the inclusive signal: the measurement of an enhancement at the LHC may be a sign of unexpected new physics.
Note also that some very different models can give the same prediction, like the fourth generation case where a suppression in the $\gamma \gamma$ decay is accidentally compensated by an enhancement in the gluon fusion cross section.
Therefore, we need to measure another observable at the LHC in order to distinguish such models.
For the light Higgs case, in Figure~\ref{fig:light} we plotted the vector boson fusion channel, which is sensitive to the $\gamma \gamma$ branching fraction directly.
This channel is orthogonal to the  inclusive one, and therefore offers the best discrimination power.
Experimentally, there is a hope to measure this channel with a very large luminosity~\cite{VBFgamma}.
For a heavier Higgs, in Figure~\ref{fig:heavy}, the decay in massive gauge bosons $H \to V^* V$ (with one virtual) becomes relevant and offers another discovery channel.
This channel, sensitive to the total cross section, will allow for a discrimination for Higgs masses near the $WW$ threshold.

\begin{figure}[!t]
\begin{center}
\includegraphics[width=1\textwidth]{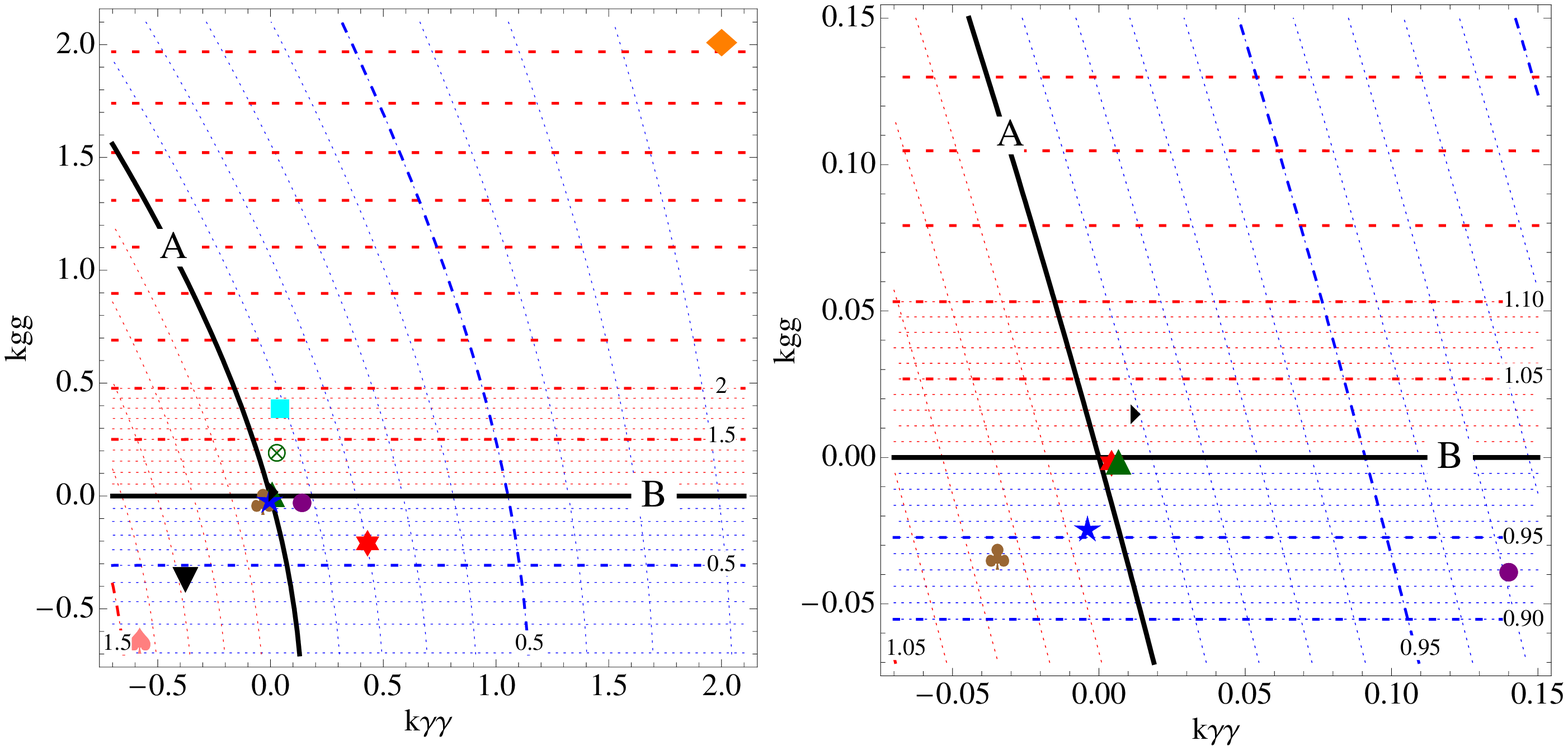}
\end{center}
\caption{\footnotesize  $\kappagamma$ and $\kappaglu$ at the ILC ($m_H = 120$ GeV).
The two solid lines correspond to the SM values of the $\gamma \gamma$ ({\bf A}) and gluon ({\bf B}) branching ratios.
On the left panel, the dashed lines are spaced by $0.5$, while the dotted ones by $0.1$. On the right, we zoomed near the SM point.
}
\label{fig:ILC}
\end{figure}

The Linear Collider will have a much better chance to discriminate between models than the LHC.
In fact, an experiment at a linear collider will be able to measure directly the branching fractions into gluons and photons.
After 100 fb$^{-1}$ of data, in the photon channel an accuracy of 5--7 \% is expected (reduced to 2--3 \% with the $\gamma \gamma$ collider option), while the gluon channel offers a 2 \% accuracy (assuming SM values)~\cite{ILCreach}.
We compared the models with the ILC measurements in Figure~\ref{fig:ILC}.

\section{Conclusions}
\label{sec:conclusion}
\setcounter{equation}{0}

The decay in a pair of photons is the golden channel for the discovery of the Higgs boson at the LHC for an intermediate mass, below the $WW$ threshold, where the dominant decay mode would be $\bar{b} b$.
This decay occurs via a loop diagram, where the heaviest particles in the SM ($W$ and top) contribute the most.
Furthermore, the production cross section at the LHC is dominated by a similar loop diagram that mediates the coupling of the Higgs to a pair of gluons.
This situation offers a precious handle on new physics: in fact, new particles that may be present at the TeV scale will also contribute to those loops, therefore modifying the SM predictions for the Higgs production and decay rates.

One of the main motivations to expect new physics at the TeV scale is the naturalness of the Higgs mass (electroweak scale): the new particles, partners of the gauge bosons and of the top, will cancel or soften the divergences in the loop corrections to the Higgs mass.
If this is the case, the new particles will have a significant coupling to the Higgs and therefore contribute significantly to the loop couplings of the Higgs.
The LHC will be able to discover such new particles, with masses up to few TeV for particles with strong interactions and 1 TeV for weakly interacting ones.
However, little information on the couplings will be directly accessible: the discovery of new states will not tell us if they play any role in the Higgs physics.
Measuring deviations in the $H \to \gamma \gamma$ and $H \to gg$ couplings at later times will give us an important hint to understand the nature of the new states and of the underlying model of electroweak symmetry breaking.

In this paper, we studied the contribution of new physics to the $H \to \gamma \gamma$ and $H \to gg$ decay widths (the latter is proportional to the production cross section).
We propose a convenient parameterisation of the new contributions, by introducing two independent parameters $\kappagamma$ and $\kappaglu$.
Simple new physics scenarios give rise to simple correlations in this parameter space: for instance, a top partner will have $\kappagamma = \kappaglu$, while a single new particle will generate same-sign $\kappa$'s.
In order to illustrate the power of a model independent measurement at the LHC (and at future Linear Colliders) we compiled a necessarily incomplete survey of models of new physics both in 4 and 5 dimensions.
Our results show that there are classes of models pointing in different quadrants of the parameter space, and that the deviations from the SM predictions can be as large as 50\%.
Moreover, in most cases those results do not depend on the details of the model and they are sensitive to just one mass scale of the new physics.
Therefore, a cross section measurement at the LHC will allow to discriminate models even with new particle masses at the TeV scale.
At the Linear Collider, the few percent level measurement of the Higgs branching ratios will allow an even better discrimination.
Note also that most of the models in our survey populate the $\kappagamma < \kappaglu$ region, where we generically expect a suppression of the inclusive cross section.
In this parameterisation it would be easy to discover hints of unconventional or unexpected new physics, independently on direct and/or indirect signals in other channels.

\section*{Acknowledgements}
We thank Suzanne Gascon and Nicolas Chanon for useful discussions and comments.
This work is supported in part by the ANR project SUSYPHENO (ANR-06-JCJC-0038).

\newpage
\appendix

\section{Appendix: Higgs couplings in extended Higgs sectors}
\setcounter{equation}{0}
\label{app:general}

\subsection{Multiple Higgs}

The Higgs sector may contain multiple scalar fields which develop a VEV, like for instance in supersymmetry where two Higgs doublets are required in order to allow up and down type Yukawa interactions.
Let us imagine that there are $n$ such Higgs multiplets $\phi_i$, such that
\beq
\phi_i = \frac{1}{\sqrt{2}} \left( v_i + c_i h + \dots \right)\,,
\eeq
where $h$ is the lightest mass eigenstate (that we would identify with the SM Higgs), and dots represent the other (heavier) scalar mass eigenstates.
The VEVs are all non zero $v_i \neq 0$.
In this case, in the formulae in Section~\ref{sec:definitions}, one needs to replace:
\beq
\frac{v}{m} \frac{\partial m}{\partial v} \to  \frac{v}{m} \sum_i  \frac{\partial m}{\partial v_i} c_i\,.
\eeq

For example, in the case of supersymmetry, there are two Higgs doublets, $H_{u,d}$ with $H_u = 1/\sqrt{2} ( v_u + h \cos \alpha + \sin\alpha H)$ and $H_d = 1/\sqrt{2} ( v_d - h \sin \alpha + \cos\alpha H)$, and $v^2 = v_u^2 + v_d^2$ ($\tan \beta = v_u/v_d$).
The $W$ mass is given by $m_W^2 = g^2/4 (v_u^2 + v_d^2)$, so that:
\beq
\frac{v}{m_W} \frac{\partial m_W}{\partial v} \to \frac{v}{m_W} \left( \frac{m_W}{v} \sin \beta \cos \alpha - \frac{m_W}{v} \cos \beta \sin \alpha \right) = \sin (\beta - \alpha)\,.
\eeq
For the top, $m_t = y v_u$:
\beq
\frac{v}{m_t} \frac{\partial m_t}{\partial v} \to \frac{v}{m_t} y \cos \alpha = \frac{\cos \alpha}{\sin \beta}\,.
\eeq

\subsection{Higgs mixing}

Another interesting case is when the Higgs mixes with additional scalars that do not develop a VEV.
This situation may be realised in multiple Higgs models, or in the Lee-Wick SM.
We will call $S_j$ those inert scalars, which contain the light Higgs field h:
\beq
S_j = s_j h + \dots
\eeq
As before, we are assuming that the other mass eigenstates are heavier than the $h$, which we want to identify with the SM Higgs.
The scalars $S_j$ may couple to a particle $p$ with coupling
\beq
g_{S_j}\; S_j\, \bar p p\,,
\eeq
which will contribute to the coupling of $p$ to the Higgs $h$ via the mixing.
In this case, one can use the formulae in Section~\ref{sec:definitions} with
 \beq
\frac{v}{m} \frac{\partial m}{\partial v} \to  \frac{v}{m} \left(\sum_i  \frac{\partial m}{\partial v_i} c_i + \sum_j g_{S_j} s_j \right) \,.
\eeq

\subsection{Charged Higgs couplings} \label{app:GB}

Another situation where the coupling to the Higgs does not come via the $v$-dependence of the mass, is when the particle in question does couple with the Higgs potential.
In fact, the Higgs potential implicitly contains the VEV, and this fact may lead to cancellations in the particle mass.
One may calculate the mass of the particle as a function of the Higgs field VEV $\langle H \rangle$ and $v$ (which are numerically equal), derive in  $\langle H \rangle$ and then impose $\langle H \rangle = v$.
In most cases it is easier to compute the coupling directly from the Higgs potential: here we will summarise three cases that are useful for the calculations in this paper.

The most trivial example is the charged Goldstone boson in the SM, which is eaten by the $W$ in the Unitary gauge. In Feynman gauge, the charged component of the Higgs doublet remains in the spectrum and its mass is $m_{\phi^\pm} = m_W$.
This may lead to the wrong conclusion that its couplings to the Higgs are the same as the $W$.
The Higgs potential can be written as
\beq
V(H) = \frac{\lambda}{4} \left( H^\dagger H - \frac{v^2}{2} \right)^2\,;
\eeq
after expanding the Higgs field as $H = \left( \begin{array}{c} \phi^+ \\ \frac{v+h + i\phi}{\sqrt{2}} \end{array} \right)$,  it does not generate any mass for the Goldstone bosons $\phi$, because of a cancellation between the mass term $- \lambda v^2/4$ and a contribution from the quartic coupling (the mass is given by the gauge fixing term).
However, the quartic coupling does generate a trilinear coupling with the Higgs $h$:
\beq
\frac{\lambda v}{2} \phi^+ \phi^- h = \frac{m_h^2}{v} \phi^+ \phi^- h\,.
\eeq
The coupling to the Higgs is therefore proportional to the Higgs mass.
The amplitude generated by the Goldstone boson can be computed starting from the amplitude of a standard scalar 
\beq
A_{\phi^\pm} (\tau_W)= \frac{v}{2 m_W^2} \frac{m_h^2}{v} A_S (\tau_W) = 2 \tau_W A_S (\tau_W)\,,
\eeq
where $\tau_W = \frac{m_h^2}{4 m_W^2}$.

A similar situation happens in the Lee-Wick SM: together with the standard Higgs field $H$, there exist a LW scalar $\tilde{H}$ with negative kinetic term.
The potential is:
\beq
V_{LW} (H,\tilde H) = V (H-\tilde H) - M_H^2 \tilde{H}^\dagger \tilde{H}\,.
\eeq 
Only the standard Higgs develops a VEV, while the LW Higgs does not thanks to its large LW mass $M_H$.
The charged component of the Higgs is eaten by the massive $W$; the charged component of the LW field $\tilde{h}^+$ is a physical degree of freedom with mass given simply by the LW mass: the $v$ dependence cancels out like for the Goldstone bosons.
Nevertheless, a trilinear coupling $\tilde{h}^+ \tilde{h}^- h$ is present with coefficient proportional to $\lambda v/2$ (the proportionality coefficient depends on the mixing in the neutral sector, and it is discussed in Section~\ref{sec:models4}).
The amplitude for the LW field can be written as:
\beq
A_{\tilde{h}^\pm} (\tilde{\tau}_{h^\pm}) =  \frac{v}{2 \tilde{m}_{h^\pm}^2} \frac{\lambda v}{2} A_S (\tilde{\tau}_h) =  \frac{\lambda v^2}{2 m_h^2} \,  2 \tilde{\tau}_h A_S (\tilde{\tau}_h)\,;
\eeq
where $\tilde{\tau}_h = \frac{m_h^2}{4 \tilde{m}_{h^\pm}^2}$.
This formula is different from the one used in Ref.~\cite{Krauss:2007bz}.

Finally, let us discuss the case of the charged Higgs in the MSSM: the model contains two Higgs doublets $H_{u,d}$ with opposite hypercharge to generate up- and down-type Yukawas.
The potential contains two quartic couplings~\cite{susyprimer}:
\begin{multline}
V_{MSSM} (H_u, H_d) = \frac{g^2 + {g'}^2}{8} \left( |H_u|^2 - |H_d|^2 \right)^2 + \frac{g^2}{2} \left| H_u H_d^\dagger \right|^2 + \dots \\
   =   \frac{g^2 + {g'}^2}{8} \left( |H_u^0|^2 - |H_d^0|^2 + H_u^- H_u^+ - H_d^+ H_d^- \right)^2 + \frac{g^2}{2} \left| H_u^+ (H_d^0)^* + H_u^0 H_d^+ \right|^2 + \dots
  \end{multline}
  where the dots stand for quadratic terms.
The two neutral components develop a VEV: $\sqrt{2}\, \langle H_u^0 \rangle = v_u = v \sin \beta$ and $\sqrt{2}\, \langle H_d^0 \rangle = v_d = v \cos \beta$.
However, only one combination actually acquires a VEV: we can define $H_1 = \sin \beta\, H_u - \cos \beta\, H_d^\dagger$ and $H_2 = \cos \beta \, H_u + \sin \beta\, H_d^\dagger$ such that $\sqrt{2}\, \langle H_1 \rangle = v$ and $\sqrt{2}\, \langle H_2 \rangle = 0$.
The charged component of $H_1$ is eaten by the $W$, while the charged component of $H_2$ is the physical charged Higgs: $H_u^+ = \cos \beta\, H^+$ and $H_d^+ = \sin \beta\, H^+$.
Plugging those solutions in the potential, and expanding around the VEV $\sqrt{2}\, \langle H_u^0 \rangle = v_u + \cos \alpha\, h + \sin \alpha H$ and $\sqrt{2}\, \langle H_d^0 \rangle = v_d - \sin \alpha\, h + \cos \alpha H$, we find that
\beq
m_{H^\pm}^2 &=& m_A^2 + m_W^2 \,, \\
g_{H^+ H^- h} &=& \frac{2 m_W^2}{v} \sin (\beta - \alpha) + \frac{m_Z^2}{v} \cos (2 \beta) \sin (\beta + \alpha)\,,
\eeq
where $m_A$ is a mass term independent on the VEVs.
The coupling to the light Higgs has a term proportional to the $W$ mass square, coming from the second term in the potential (this is what we would obtain from the mass formula), and a term proportional to the $Z$ mass square, from the first quartic term in the potential: the latter cancels out in the mass formula but does contribute to the Higgs couplings.

\section{Appendix: gauge bosons in 5D}
\setcounter{equation}{0}
 
In this appendix, we propose a more detailed description of the models that we consider in section~\ref{sec:models5}, we sketch how to extract the spectra of masses for Gauge bosons for different choices of geometry and compactification of the fifth dimension. 
We first derive general results in a generic metric with the extra coordinate $y\, \epsilon\; [y_1, y_2]$ and
\beq
d s^2 = w (y)^2 ( d x_\mu d x^\mu - d y^2)\,,
\eeq
and then we discuss the limits of flat ($w = 1$, $y_1=0$ and $y_2=\pi L$) and warped AdS ($w = R/y$, $y_1=R$ and $y_2=R'$) metrics.
The action of a pure gauge theory in one extra dimension, after fixing the $R_\xi$ gauge, is given by: 
\begin{equation}
\mathcal{S} = \int \ud^4x \int_{y_1}^{y_2} \ud y\, w\, \left\{ -\frac{1}{4} F^a_{\mu\nu} F^{\mu\nu a}-\frac{1}{2}F^a_{\mu5} F^{\mu5a} - \frac{1}{2 \xi}\left[\pd_\mu A^{\mu a} - \xi \frac{1}{w}  \pd_5 \left( w A_5^{a} \right) \right]^2 \right\}\,, \label{action2}
\end{equation}
where $F^a_{MN}=\pd_M A^a_N-\pd_N A^a_M+g_5f^{abc}A^b_MA^c_N$ and $g_5$ is the 5D gauge coupling.
In the Unitary gauge ($\xi \to \infty$), the massive modes of the fifth component $A_5$ are removed and they become the longitudinal polarisation of the massive KK vectors.
In the following, we will discuss two models: one where the Higgs is part of the gauge field, namely a zero mode of the $A_5$, and another where the Higgs is a brane localised field.

\subsection{Gauge Higgs Unification Models}
\label{app:gaugehiggs}

A zero mode for the $A_5$ component of the gauge field is a physical scalar in the spectrum because it is not eaten up in the Unitary gauge.
However, it is a special scalar because its potential is constrained by Lorentz and gauge invariance: in 5D no potential is allowed at tree level, therefore it is generated at loop level and it is finite.
This property makes the $A_5$ an ideal candidate to play the role of the Higgs boson.
In order to obtain a zero mode, we need to enlarge the SM gauge group such that a doublet of SU(2) is part of the gauge fields, and break the gauge directions of this doublet on both end points by imposing Dirichlet boundary conditions on the vectors (and therefore Neumann boundary conditions on the $A_5$ component).

For simplicity, we work on the minimal model where the gauge symmetry is enlarged to $SU(3)$, broken to the electroweak $SU(2) \times U(1)$ at the boundaries. 
The bulk fields can be written as:
\begin{align}
\ A_\mu=&\left[\begin{array}{ccc}
\scriptstyle{\boldsymbol{W^{(3)}_\mu}+\boldsymbol{1/\sqrt{3}B^{(8)}_\mu}} & \boldsymbol{W^+_\mu} & D^+_\mu \\
\boldsymbol{W^-_\mu} & \scriptstyle{-\boldsymbol{W^{(3)}_\mu}+\boldsymbol{1/\sqrt{3}B^{(8)}_\mu}} & D^0_\mu \\
D^-_\mu & D^{0\dagger}_\mu & \scriptstyle{\boldsymbol{2/\sqrt{3}{B^{(8)}_\mu}}}\\
\end{array}\right] 
\ \mbox{and} \ \ A_5=\left[\begin{array}{ccc}
0& 0 & \boldsymbol{H^+}  \\
0 & 0 &  \boldsymbol{H^0} \\
 \boldsymbol{H^-} &  \boldsymbol{H^{0\dagger}} & 0\\
\end{array}\right] 
\label{gaugefield}
\end{align}
where $W$ and $B$ are towers with a zero mode, $D$ are massive gauge bosons and $H$ is the Higgs field (only the zero mode).
We assume that the radiative potential will generate a VEV for the Higgs
\beq
\langle H^0 \rangle = \frac{V}{\sqrt{2}} \frac{1}{w(y)}\,,
\eeq
where $V$ is a constant and the $y$ dependence is encoded in the metric factor $w$.
The presence of this VEV will affect the bulk equation of motions for all fields: however, being $H$ part of gauge fields, we can use an SU(3) gauge transformation to remove the VEV from the bulk equation of motions, and cast it into the boundary conditions~\cite{gaugetransfo}. 
For the gauge bosons, we can define:
\begin{align}
\tilde {A}_M &= \Omega(y)  A_M  \Omega^\dagger(y) -  \frac{i}{g_5} \Omega(y)  \pd_M  \Omega^\dagger(y) \nonumber \\
\mbox{so that} \quad \langle \tilde  {A}_5\rangle &= \Omega(y) \langle A_5\rangle \Omega^\dagger(y)  - \frac{i}{g_5} \Omega(y)  \pd_y  \Omega^\dagger(y) =0\,.
\end{align}
The gauge transformation that does this job can be written as:
\beq
\Omega(y) = \exp \left[ i g_5 v/\sqrt{2} \int_{y_1}^y dy' \frac{1}{w (y')}  \lambda_7 \right]\,,
\eeq
where $\lambda_7$ is the generator of SU(3) aligned with $H^0$.
Note that we fixed the gauge transformation such that it only affects one brane: in fact $\Omega (y_1) = 1$, and
\beq
\Omega (y_2) = \exp \left[ i \pi \alpha \lambda_7 \right] = \left( \begin{array}{ccc}
                    1 & 0 & 0\\
0& \cos\pi\alpha& i\ \sin\pi\alpha\\
0& i\ \sin\pi\alpha & \cos\pi\alpha
\end{array}\right) \,,
\eeq
where
\beq
\alpha = \frac{g_5 V}{\sqrt{2}} \int_{y_1}^{y_2} \frac{dy}{\pi} \frac{1}{w (y)}
\eeq
is a dimensionless parameter proportional to the Higgs VEV $V$.
The equations of motion of the new fields do not depend on the Higgs VEV, however the boundary conditions on one end will be affected.
For example, for the charged gauge bosons, the gauge transformation will mix $W^+$ and $D^+$, which have respectively Neumann and Dirichlet boundary conditions on both endpoints: in the new basis
\begin{eqnarray}
\left\{ \begin{array}{c} 
D^+_\mu(y_1)=\tilde{D}^+_\mu(y_1)= 0  \\
\pd_5 W^+_\mu(y_1)=\pd_5\tilde{W}^+_\mu(y_1)= 0
\end{array} \right.\quad \mbox{and}\quad
\left\{ \begin{array}{c}
D^+_\mu (y_2)=\cos \pi \alpha\, \tilde{D}^+_\mu  + i \sin \pi \alpha\, \tilde{D}^+_\mu  = 0 \\
\pd_5 W^+_\mu(y_2)=\cos \pi \alpha\, \pd_5 \tilde{W}^+_\mu + i \sin \pi \alpha\, \pd_5 \tilde{D}^+_\mu = 0
\end{array} \right.\nonumber
\end{eqnarray}

The specific form of the spectrum depends on the metric: in the flat case
\beq
\alpha = \frac{g_5 V}{\sqrt{2}} \int_{0}^{\pi L} \frac{dy}{\pi} = \frac{g_5 L}{\sqrt{2}} V\,,
\eeq
and
\beq
\left\{ \begin{array}{c} \tilde{W} (x,y) \\ \tilde{D}(x,y) \end{array} \right\} = \sum_n  \left\{ \begin{array}{c} (A_n \cos m_n y + B_n \sin m_n y) \\  (C_n \cos m_n y + D_n \sin m_n y) \end{array} \right\}  W_n(x)\,.
\eeq
Applying the boundary conditions to such wave functions, we obtain that the spectrum is determined by the solutions of the following equation:
\beq
\sin\left(m_nL\pm \pi \alpha\right) =0 \,.
\eeq

The warped case is more complicated because the solutions of the equations of motion are Bessel functions of the first and second kind of order 1:
\beq
\alpha = \frac{g_5 V}{\sqrt{2}} \int_{R}^{R'} \frac{dy}{\pi} \frac{y}{R} = \frac{g_5 {R'}^2}{\pi \sqrt{2} R} V\, \left( 1 - \frac{R^2}{{R'}^2} \right)\,,
\eeq
and
\beq
\left\{ \begin{array}{c} \tilde{W} (x,y) \\ \tilde{D}(x,y) \end{array} \right\} = \sum_n  \left\{ \begin{array}{c} y (A_n J_1(m_n y) + B_n Y_1 (m_n y)) \\ y  (C_n J_1 (m_n y) + D_n Y_1 (m_n y)) \end{array} \right\}  W_n(x)\,.
\eeq


\subsection{Brane Higgs Models}
\label{app:braneHiggs}

In these models, the Higgs boson is a 4D field which couples with the 5D gauge bulk field only on a boundary, so that the Higgs VEV only enters in the boundary conditions.
We will first focus on the case where the bulk gauge symmetry is the same as in the SM, without extra fields that mix with the $W$: the action in the bulk is the same as in \eqref{action2} and the 5D field can be KK decomposed as we have done before.
The boundary conditions on the two endpoints can be written as (here we assume the Higgs localised on $y_2$, but the results do not depend on this choice)
\beq
\left\{ \begin{array}{c}
\pd_5W^+_\mu(y_1)= 0\\
\pd_5 W^+_\mu(y_2)+\frac{g_5^2 V^2}{4}\ W^+_\mu(y_2) = 0\end{array} \right.
\eeq
where $V$ is the Higgs VEV and $g_5$ the 5D gauge coupling.
If we decompose the 5D fields as usual
\begin{eqnarray}
W^+ (y, x) = \sum_n f_n(m_n y) \ W_n(x)^+\,,
\end{eqnarray}
the second boundary condition determines the spectrum as the solutions of the equation
\beq
m_n \frac{f' (m_n y_2)}{f (m_n y_2)} + \frac{g_5^2 V^2}{4} = 0\,. \label{eq:spectrumbH}
\eeq
The precise form of the function $f$ depends on the geometry after imposing the boundary condition on the other endpoint.

In the flat case
\beq
f (m_n y) = \sin (m_n y) \Rightarrow \pi L m_n \tan \pi L m_n - \pi^2 \alpha^2 = 0\,,
\eeq
where we have defined for convenience
\beq
\alpha = \sqrt{\frac{L}{\pi}} \frac{g_5 V}{2}\,.
\eeq
In the warped case
\beq
f (m_n y) = y ( Y_0 (m_n R) J_1 (m_n y) - J_0 (m_n R) Y_1 (m_n y) )\,.
\eeq

Note finally that the simple form of eq.~(\ref{eq:spectrumbH}) allows us to calculate the couplings of the n-th mode to the Higgs as a function of the mass, even though the mass cannot be explicitly calculated: in fact, taking the total derivative with respect to $V$ and eliminating $V$ by using eq.~(\ref{eq:spectrumbH}), we obtain
\beq
\frac{v}{m_n} \frac{\partial m_n}{\partial v} = \frac{2 \frac{f'}{f}}{\frac{f'}{f} + m_n y_2 \left( \frac{f''}{f} - \left( \frac{f'}{f} \right)^2 \right)}\,.
\eeq
By studying this expression numerically or in an expansion for small $\alpha$, we found that the sum rule
\beq
\sum_n \frac{v}{m_n} \frac{\partial m_n}{\partial v} = 1\,,
\eeq
where we are summing over all the mass eigenstates, is respected both in the flat and warped case.


\section{Appendix: Fermionic fields}
\label{app:fermions}
\setcounter{equation}{0}

Here we are considering the minimal 5D bulk action for a fermionic field $\Psi$:
\begin{equation}
\mathcal{S} = \int \ud^4x  \int_{y_1}^{y_2} \ud y\,  w (y)^4 \; \left[ \frac{i}{2}\left( \overline{\Psi} \Gamma^M \pd_M\Psi - \pd_M\overline{\Psi} \Gamma^M  \Psi      \right)- w(y) \tilde{M} \overline{\Psi} \Psi \right] \label{action}
\end{equation}
where $\Gamma^M$ with $M=1 \ldots 5$ are the five Dirac $4\times4$ matrices for 5D representation of the Clifford Algebra and $\tilde{M}$ is the odd bulk mass of the 5D fermion. We remind that in 5D, the irreducible Lorentz representation for $\Psi$ is a Dirac spinor, which is not chiral.
For convenience we can use the Weyl spinor notation $\Psi= \left(\begin{array}{c}
\chi\\\bar{\psi}\end{array}\right)$; thus, the equations of motion for the fermionic bulk fields are given by:
\begin{eqnarray}
\left\{ \begin{array}{c}
-i\bar{\sigma}^\mu \pd_\mu \chi - \pd_5\bar{\psi}+ \left( w\tilde{M} - 2 \frac{w'}{w} \right)\bar{\psi} = 0  \\ 
-i\sigma^\mu \pd_\mu \bar{\psi} + \pd_5 \chi+\left( w\tilde{M}+ 2 \frac{w'}{w} \right) \chi = 0\end{array} \right.
\end{eqnarray}
The next step consists in using the KK decomposition of the 5D spinors to extract the evolution along the fifth dimension. The components $\chi$ and $\psi$ are defined by:
\begin{equation}
\chi(x,y) = \sum_{n}g_n(y) \chi_n(x) \quad \mbox{and} \quad
\bar{\psi}(x,y)= \sum_{n} f_n(y) \bar{\psi}_n(x)\,, 
\end{equation}
where $\chi_n(x)$ and $\bar{\psi}_n(x)$ are the two 4D-components of the Dirac field with the mass $m_n$ and satisfying usual 4D Dirac equations
\beq
\left\{\begin{array}{c}
-i\bar{\sigma}^\mu \pd_\mu \chi^{(n)} +m_n\bar{\psi^{(n)}} = 0\\
-i\sigma^\mu \pd_\mu \bar{\psi^{(n)}} +m_n \chi^{(n)} = 0
\end{array}\right.
\eeq
The wave functions therefore will satisfy the following equations
\beq
\mbox{flat case:} &\Longrightarrow&
\left\{\begin{array}{c}
g'_n+{\tilde{M}}g_n-m_nf_n=0 \\
f'_n-{\tilde{M}}f_n+m_ng_n=0
\end{array}\right. \\
\mbox{AdS case:} &\Longrightarrow&
\left\{ 
\begin{array}{c}
 g'_n+\frac{c-2}{y}g_n-m_nf_n=0 \\
f'_n-\frac{c+2}{y}f_n+m_ng_n=0
\end{array}\right.
\eeq
where we have defined $c=\tilde{M} R$ in the warped case.
The solutions of those equations in the flat case will be combinations of $\sin (\sqrt{m_n^2 - \tilde{M}^2} y)$ and $\cos (\sqrt{m_n^2 - \tilde{M}^2} y)$ (which become hyperbolic for the massless/light mode).
In the warped case we have Bessel functions $J_{1/2 \pm c} (m_n y)$ and $J_{-1/2 \pm c} (m_n y)$ ~\cite{fermions}.

 Finally we have to consider models with chiral SM fermions. This is achieved by taking boundary conditions for the Dirac fields which allow light chiral zero modes:
\begin{eqnarray}
\left. \begin{array}{c}
\mbox{Left-handed} \\
\mbox{fermion}
\end{array} \right. \rightarrow 
\psi\mid_{y_1,y_2} = 0
\left\| \begin{array}{c}
\\
\\
\end{array} \right. \left. \begin{array}{c}
\mbox{Right-handed}\\
\mbox{fermion}
\end{array}\right.
 \rightarrow 
\chi\mid_{y_1,y_2} = 0\label{BC's}
\end{eqnarray}
To complete the description of fermions and to relate it to SM phenomenology, we need to introduce two bulk fields, a singlet $\Psi_R$ with a right-handed zero mode and a doublet of $SU(2)$ $\Psi_L$ with a left-handed zero mode, and their couplings with the Higgs boson. From here, we need to specify some properties of the 5D models.

\subsection{Gauge Higgs Unification Models}

In this case, the singlet $\Psi_R$  and the doublet $\Psi_L$  are embedded in the same bulk field, a representation of the larger bulk gauge symmetry.
Consequently the odd bulk mass $\tilde{M}$ is the same for the doublet and the singlet components.
The interaction with the Higgs boson appears in the covariant derivative of $\Psi$ in the kinetic term. 
This additional term in the action is given by $-ig_5\bar{\Psi} \Gamma^5 A_5\Psi$ in the bulk: the bulk Yukawa coupling is therefore proportional to the gauge coupling $g_5$ and the proportionality factor depends on the specific representation of $\Psi$.
This term will modify the bulk equations of motion: however, as in the gauge boson case, we can use a gauge transformation to remove the Higgs VEV, and recast its effects on one of the boundary conditions.

Here we will focus on the SU(3) case described in the text for simplicity.
The gauge transformed fields on the $y_2$ brane are
\beq
\tilde{\Psi}(y_2)= \Omega_f(y_2) \Psi(y_2) \qquad
\mbox{where} \quad  \Omega_f(y_2) =  \exp \left[ i\ \pi \alpha \tilde{\lambda}_7 \right]\,, 
\label{redef}
\eeq
where $\tilde{\lambda}_7$ is the SU(3) generator in the representation of $\Psi$.
The matrix $\Omega_f$ will mix the singlet and the component of the doublet which picks up a mass (for simplicity we will denote it with $\Psi_L$).
The mixing angle however, is not $\alpha$ in general: in fact it will depend on the representation of the bulk field, and the proportionality factor can be calculated by explicitly computing the generator $\tilde \lambda_7$ for the bulk fermion representation.
In general, we will define a new parameter $\beta$ to describe the mixing.
Note that in the case of a bulk fundamental, $\Omega$ is the same as the one used for the gauge bosons in the previous section, therefore $\beta ({\bf 3 })= \alpha$. 
The new boundary conditions for the transformed fields are
\beq
\left\{ \begin{array}{c}
\psi_L (y_2) = \cos \pi \beta \, \tilde{\psi}_L (y_2) - i \sin \pi \beta\,  \tilde{\psi}_R (y_2)= 0 \\
\chi_R (y_2) = \cos \pi \beta\, \tilde{\chi}_R(y_2) - i \sin \pi \beta\, \tilde{\chi}_L (y_2) = 0
\end{array} \right.
\eeq
This boundary conditions will determine the spectrum: for instance, the spectrum ${m_n}$ in the flat case is given by the solutions of
\begin{equation}
-\cos  2\pi  L \sqrt{-\tilde{M}^2 + m_n^2} + \cos 2 \pi \beta + 2 \frac{\tilde{M}^2}{m_n^2} \sin^2 \pi \beta=0\,.
\end{equation}

\subsection{Brane Yukawas}

Fermionic masses can also be generated by Yukawa couplings localised on an endpoint of the extra dimension: this is possible both in the bulk Higgs model and in the localised Higgs case.
Like in the Gauge Higgs case, the Higgs VEV only enters in the boundary conditions. 
However, boundary conditions for fermions are more tricky than for bosons, due to the fact that the equations of motion are first order differential equations: therefore how the VEV enters the boundary conditions depends crucially on the localisation mechanism for the Higgs field or for the Yukawa couplings (see Ref.~\cite{fermions}).
Here we will consider the simplest possibility: that the boundary conditions are linear in the Higgs VEV:
\beq
\left\{ \begin{array}{c}
\psi_L - y V L\psi_R = 0 \\
\chi_R + y V L\chi_L = 0
\end{array} \right.
\eeq
Those boundary conditions are the same as in the gauge Higgs case if we identify $\tan \pi \beta =  y V L $ (and removing the $i$ with a phase redefinition of the fields).
The only difference is that $\beta$ is not proportional to the Higgs VEV, therefore additional corrections to the couplings will arise.
Another novelty is that the singlet and doublet fields are part of different bulk fields, therefore they can have different bulk masses $\tilde{M_L}$ and $\tilde{M_R}$.


\end{document}